\documentclass[11pt]{article}
\usepackage[utf8]{inputenc}

\title{Too Noisy to Collude? Algorithmic Collusion Under Laplacian~Noise}
\author{
Niuniu Zhang\footnote{UCLA Anderson School of Management (\href{mailto:niuniu.zhang.phd@anderson.ucla.edu}{niuniu.zhang.phd@anderson.ucla.edu}). Preliminary draft, written as my first-year PhD summer paper. I thank Auyon Siddiq and colleagues for helpful feedback. Comments welcome.}
}

\usepackage{amsmath,amssymb,amsthm}

\usepackage{tikz}
\usetikzlibrary{calc}
\usepackage[margin=1in]{geometry}
\usepackage{xcolor}
\usepackage{color}
\usepackage{xspace}
\usepackage{algorithm,algpseudocode}
\usepackage{bm}
\usepackage[hyperfootnotes=false, colorlinks=true, linkcolor=black, urlcolor=black, citecolor={blue}]{hyperref}
\usepackage{soul}
\usepackage{framed}
\usepackage{caption}
\usepackage{subcaption}
\usepackage{tikz}
\usepackage{tikz-cd}
\usepackage{enumerate}
\usepackage{listings}
\usepackage{booktabs}
\usepackage{physics}

\colorlet{punct}{red!60!black}
\definecolor{background}{HTML}{EEEEEE}
\definecolor{delim}{RGB}{20,105,176}
\colorlet{numb}{magenta!60!black}

\lstdefinelanguage{json}{
    basicstyle=\normalfont\ttfamily,
    numbers=left,
    numberstyle=\scriptsize,
    stepnumber=1,
    numbersep=8pt,
    showstringspaces=false,
    breaklines=true,
    frame=lines,
    backgroundcolor=\color{background},
    literate=
     *{0}{{{\color{numb}0}}}{1}
      {1}{{{\color{numb}1}}}{1}
      {2}{{{\color{numb}2}}}{1}
      {3}{{{\color{numb}3}}}{1}
      {4}{{{\color{numb}4}}}{1}
      {5}{{{\color{numb}5}}}{1}
      {6}{{{\color{numb}6}}}{1}
      {7}{{{\color{numb}7}}}{1}
      {8}{{{\color{numb}8}}}{1}
      {9}{{{\color{numb}9}}}{1}
      {:}{{{\color{punct}{:}}}}{1}
      {,}{{{\color{punct}{,}}}}{1}
      {\{}{{{\color{delim}{\{}}}}{1}
      {\}}{{{\color{delim}{\}}}}}{1}
      {[}{{{\color{delim}{[}}}}{1}
      {]}{{{\color{delim}{]}}}}{1},
}

\usepackage{bbm}

\usepackage{setspace}
\newcommand{\inparen}[1]{\left( #1 \right) }
\newcommand{\inbrak}[1]{\left[ #1 \right] }

\newcommand{\inset}[1]{\left\{ #1 \right\} }

\newcommand{\argmax}{\operatorname{argmax}}

\newcommand{\p}{\mathbb{P}}

\newcommand{\BNE}{\operatorname{BNE}}

\newcommand{\E}{\mathbb{E}}

\newtheorem{lemma}{Lemma}
\newtheorem{theorem}{Theorem}

\newtheorem{definition}{Definition}

\newtheorem{proposition}{Proposition}

\newcounter{protocol}

\usepackage{natbib}
 \bibpunct[, ]{(}{)}{,}{a}{}{,}%

\begin{document}

\date{September 1, 2025}
\maketitle

\begin{abstract}
\begin{spacing}{1.}
The rise of autonomous pricing systems has sparked growing concern over algorithmic collusion in markets from retail to housing. This paper examines controlled information quality as an ex ante policy lever: by reducing the fidelity of data that pricing algorithms draw on, regulators can frustrate collusion before supracompetitive prices emerge. We show, first, that information quality is the central driver of competitive outcomes, shaping prices, profits, and consumer welfare. Second, we demonstrate that collusion can be slowed or destabilized by injecting carefully calibrated noise into pooled market data, yielding a feasibility region where intervention disrupts cartels without undermining legitimate pricing. Together, these results highlight information control as a lightweight yet practical lever to blunt digital collusion at its source.
\end{spacing}
\end{abstract}

\section{Introduction}

Algorithmic pricing has raised serious policy concerns because systems that adjust prices in real time, often using pooled market data, can cause markets to move in lockstep. The danger is not explicit agreements between firms, but that reliance on the same tools can generate monopoly-like prices without communication.\footnote{Recent cases illustrate the risk: the Federal Trade Commission has alleged that Amazon secretly deployed an algorithm to raise consumer prices, and the Department of Justice (DOJ) reached a proposed settlement with Greystar, the nation’s largest landlord, over its use of RealPage’s pricing software to coordinate rents. See \href{https://www.theatlantic.com/ideas/archive/2024/08/ai-price-algorithms-realpage/679405/}{\it The Atlantic}, \href{https://www.forbes.com/sites/willskipworth/2023/10/03/amazon-allegedly-used-secret-algorithm-to-raise-prices-on-consumers-ftc-lawsuit-reveals/}{\it Forbes}, and \href{https://www.justice.gov/opa/pr/justice-department-reaches-proposed-settlement-greystar-largest-us-landlord-end-its}{\it DOJ}.} These tools may enable supracompetitive outcomes without any agreement, subverting anti-trust enforcement, which still hinges on explicit proof of communication. With legal standards lagging behind technology, we ask: \emph{how can regulators intervene in algorithmic markets to prevent collusion before it occurs?}

Pricing algorithms are no longer experimental curiosities; they are embedded across retail and rental markets because they promise efficiency and convenience for sellers. As a striking proof of concept, Anthropic recently let a large language model (LLM) manage an office shop, deciding what to stock, when to restock, and how to set prices over the course of a month.\footnote{\href{https://www.anthropic.com/research/project-vend-1}{Anthropic: ``Project Vend: Can Claude run a small shop? (And why does that matter?)"}} While this single-agent demo was not about collusion, it illustrates how quickly pricing authority is being ceded to autonomous systems. Once such agents operate in scale across firms, their adaptive updates inevitably interact. A growing body of research shows that exactly these interactions, whether through correlated learning \citep{hansen2021algorithmic, calvanoQlearning}, black-box LLMs \citep{fishLLM, keppo2025ai}, or even simple no-regret and bandit routines \citep{arun2024, douglas2024naive}, can generate cartel-like coordination without explicit communication.

Regulators face two hurdles. First, with complete pricing histories in hand, it remains notoriously difficult to separate tacit coordination from mere coincidence. Second, price patterns that strongly suggest coordination are insufficient to establish liability absent evidence of explicit communication. Under current U.S. law, no explicit communication means no conviction.\footnote{The RealPage case illustrates this gap: landlords were not penalized simply for relying on pricing software, but only once investigators uncovered evidence of direct discussions on how to set and adjust the software’s parameters for rents and strategies. See \href{https://www.justice.gov/opa/pr/justice-department-reaches-proposed-settlement-greystar-largest-us-landlord-end-its}{\it DOJ} and \href{https://www.reuters.com/legal/litigation/dc-attorney-general-inks-first-settlement-realpage-price-fixing-lawsuit-2025-06-02/}{\it Reuters}.}
Faced with these limits of legal doctrine, researchers have turned to ex post detection methods. Statistical audits or computer-science frameworks may flag suspicious behavior \citep{baranek2025detection, massarotto2025detecting}, but these tools act only after outcomes have materialized.

Inspired by differential privacy, a cryptographic protocol that adds carefully calibrated noise to data so individual records remain hidden while aggregate statistics stay accurate \citep{dwork2006dp, privacy}, we ask whether a similar principle can regulate algorithmic pricing. The idea is to blur pooled, sensitive data just enough to prevent rivals from locking into coordinated prices, yet still allow firms to obtain reliable price recommendations for legitimate business use. Crucially, this approach is ex ante: rather than waiting for coordination to emerge and relying on outdated legal thresholds, noise injection directly curbs the ability of pricing algorithms to sustain cartel-like consensus. This paper takes an information design approach, developing an analytical framework that shows how simple controls on data fidelity can meaningfully disrupt cartel coordination, highlighting a practical lever regulators could explore in future interventions. 

\subsubsection*{Model Outline}
Our starting point is a stylized model of firm competition that mirrors real-world environments such as Amazon sellers or landlords offering differentiated but substitutable products. Demand falls with a firm’s own price and rises with its rival’s, so pricing choices are inherently strategic. Firms must choose a price that maximizes their profits while anticipating that competitors are doing the same.

In practice, firms rarely observe their rivals’ demand conditions directly; such information is costly to acquire or simply private. Instead, sellers must form estimates of competitors’ market conditions. Pricing algorithms fill this gap. Because they aggregate data across many sellers, these systems see what individual firms cannot and translate high-fidelity inputs into actionable signals for pricing decisions.\footnote{For example, recent filings allege that RealPage’s software ingested sensitive, high-fidelity rent data from rival landlords and used it to generate coordinated pricing recommendations. See \href{https://www.justice.gov/archives/opa/media/1364976/dl?inline}{\it DOJ Complaint}.} In our framework, a firm enters the market, queries the algorithm for an estimate of its rival’s demand, and then sets its own price based on that signal. This setup makes clear where policy can intervene: regulators govern the quality of inputs available to pricing algorithms. By requiring algorithms to process only sanitized data with controlled fidelity, policymakers can deliberately limit the precision of the estimates that drive pricing, and thereby blunt the ability of these systems to sustain collusion.

We build the analysis in three steps:
(i) Information distortion, where firms set prices based on estimates of rival demand rather than perfect knowledge, producing biased equilibria and distortions in profit and welfare \citep{vives1984duopoly, bos2022microfoundation};
(ii) Cartel dynamics, where tacit coordination unfolds as a belief-averaging process \citep{degroot}, mediated by shared pricing algorithms rather than direct communication, and leader–follower dynamics emerge as forceful agents systematically push consensus toward higher prices \citep{acemoglu2009opinion};
(iii) Tradeoff analysis, where we first establish the incentives and harms of collusion, and then examine how much noise regulators can introduce to disrupt collusion without undermining legitimate firms’ profits or consumer welfare.

\subsubsection*{Main Results}

We contrast two benchmarks. When firms know market conditions perfectly, pricing follows standard best responses. When instead they rely on estimates of rivals’ demand, mediated by shared pricing algorithms drawing on pooled market data, outcomes immediately depend on the quality of information: prices shift away from fundamentals and the stage is set.

\textbf{Distorted competition (Information distortion)}: In Section~\ref{sec:distortion}, we quantify how imperfect information alters equilibrium outcomes: price distortions (Theorem~\ref{thm:p_dist}) that cascade into profit (Theorem~\ref{thm:pi_dist}) and consumer welfare (Theorem~\ref{thm:u_dist}). The core insight is that when pricing algorithms transmit distorted estimates, market outcomes scale directly with information quality; extreme misestimation may even price firm out of business, eliminating demand entirely. Using information quality as a regulatory lever requires careful calibration.

\textbf{How cartels form (Coordination dynamics)}: In Section~\ref{sec:cartel}, we model collusion as a social learning process over discrete rounds of negotiation \citep{degroot}, mediated by a shared pricing algorithm that all firms use.\footnote{Here, the “mediator” is simply the common pricing algorithm adopted across firms. It aggregates pooled market data and releases signals used in price setting, but it is not a communication channel between firms.} Forceful agents tilt consensus upward \citep{acemoglu2009opinion}. This forceful role naturally maps into leader-follower dynamics: the high-price firm acts as the leader, pulling the cartel toward its preferred outcome, while the low-price firm becomes the follower.

\textbf{Noise as a policy lever (Tradeoff analysis)}: In Section~\ref{sec:tradeoff}, we evaluate controlled noise as a regulatory tool. The first-order concern is validity: should collusion under our dynamics be taken seriously? We show it should. Collusion benefits leaders while harming followers. Further, contrary to common intuition, leaders’ own customers may even enjoy welfare gains when rivals’ prices rise (Theorem~\ref{thm:collusive_incentive}~and~\ref{thm:collusive_harm}). Regulators should care, though not in the way they might expect: the harm is asymmetric. The second-order concern is calibration: while noise reliably slows cartel convergence with robustness guarantee (Theorem~\ref{thm:noise_delay}, Proposition~\ref{prop:repeat_queries}), excessive noise harms legitimate firms and consumers. Our bounds trace out a feasibility region for regulators, quantifying how much noise is enough to disrupt collusion without eroding normal competition (Theorem~\ref{thm:noise_harm}).

Taken together, our results show that information quality is the central driver of outcomes. We put forward a simple but powerful idea: regulate collusion at its source by constraining what pricing algorithms can see. 

While our proposal treats information frictions in a stylized way and should be viewed as a heuristic rather than a turnkey policy, its appeal lies in being lightweight and ex ante. By targeting the data that fuels coordination, regulators gain a tractable lever to blunt cartel formation before supracompetitive prices take hold.

\section{Literature Review}
We situate our work in two strands of research: (i) evidence that algorithms can autonomously sustain collusion, and (ii) policy approaches to regulating it.

\textbf{Evidence of algorithmic collusion}: A growing body of work demonstrates that algorithms can sustain supracompetitive prices without explicit communication. \cite{calvanoQlearning} show that reinforcement-learning agents converge to collusive strategies with punishment phases. \cite{hansen2021algorithmic} demonstrate that misspecified learners, by correlating pricing experiments, drift toward monopoly outcomes, consistent with evidence from Amazon sellers. \cite{musolff2024algopricing} documents how commercial repricing tools employ “resetting” strategies that nudge rivals upward, producing cycle averages close to monopoly prices. Pushing further, \cite{douglas2024naive} prove that deterministic bandit learners always collude, while \cite{arun2024} show that no-regret learners interacting with optimizing rivals sustain high prices even without threats. 

Extending beyond traditional algorithms, \cite{fishLLM} run experiments with LLM-based pricing agents that autonomously collude, with outcomes sensitive to prompts. \cite{keppo2025ai} further show that collusion is robust under some environments but fragile under others, depending on differences in model sophistication or data access. Finally, \cite{goldstein2025} bring similar insights to finance, showing that AI trading agents collude to earn supracompetitive profits.

Most of the existing literature has focused on showing that algorithmic agents can and do collude, both in simulations and in the field. Our contribution is complementary: we model cartel formation as a belief-averaging process \citep{degroot}, enriched by forceful agents who pull outcomes toward their preferred prices \citep{acemoglu2009opinion}. This provides a social learning lens for analyzing leader–follower incentives and the asymmetric welfare consequences for consumers.

\textbf{Regulating algorithmic collusion}:
A second stream of work turns from documenting collusion to asking how it might be regulated. The leading approach is due to \cite{hartline2024algoreg,hartline2025algoregrefined}, who formalize algorithmic collusion in an online-learning framework and propose a regret-based audit that can certify plausibly non-collusive behavior. The logic is one-way: if an algorithm passes the test, it is deemed non-collusive, but failure does not by itself establish collusion. This limitation mirrors current law, where no explicit proof of communication means no conviction.

Other work develops ex post statistical tools tailored to specific environments. \cite{baranek2025detection} design a test for collusion in multistage procurement auctions, while \cite{massarotto2025detecting} draws on tools from distributed computing, such as Byzantine agreement protocols, to frame how algorithms might implicitly coordinate. Each offers valuable diagnostics, but all share the same constraint: they act only after outcomes are realized. By contrast, our study takes an ex ante perspective. We propose regulating the quality of information available to pricing algorithms themselves. Because all pricing systems ultimately rely on data inputs, this lever applies regardless of setting. Our framework shows how controlled information frictions can blunt coordination before supracompetitive prices emerge, moving from detection after the fact to prevention at the source.

Broadly speaking, our work aligns with recent research on consumer privacy, where calibrated noise in data serves as a tool to balance competing interests \citep{fainmesser2023digitalprivacy, lei2024consumerprivacy, selvi2025dpop}. Here we extend that logic from protecting individuals to restraining collusion, showing how calibrated noise in shared data can serve as a preventative remedy.

\section{Model Overview}\label{sec:linear_demand}
In this section, we introduce our general model of firm competition that underlies all our subsequent analysis.

We consider a static market with $N$ competing firms, denoted as $\inset{F_i}_{i\leq N}$. Each firm produces differentiated, but substitutable product items for a unit cost of $k_i \geq 0$, then they simultaneously set a unit price $p_i\geq 0$. Demand $D_i$ for each firm is linear, decreasing in its own price $p_i$ and increasing in the competitor's price $p_j$. Firms are strategic and myopic: each aims to maximize its own profit $\pi_i$ in a one-shot interaction, without learning or memory. Consumers respond passively to posted prices. This environment admits a tractable Nash equilibrium with closed-form pricing and consumer welfare outcomes.

To make this concrete, we focus on a duopoly linear demand setup \citep{vives1984duopoly, bos2022microfoundation}, where two firms, $F_1$ and $F_2$, compete by setting prices for their respective products. Demand for each firm is given by:
\begin{align}
    D_1 (p_1, p_2) &= a_1 - b_1 \cdot p_1 + c\cdot p_2 \\
    D_2 (p_1, p_2) &= a_2 - b_2 \cdot p_2 + c\cdot p_1 
\end{align}
Here, $a_i > 0$ represents firm $F_i$'s baseline demand, $b_i > 0$ is the sensitivity to its own price, and $c \in (0, b_i)$ is the cross-price sensitivity, capturing how demand for firm $F_i$ responds to the competitor’s price. For generality, we often write $i, j \in \{1,2\}$ with $i \neq j$.

Our main interest lies in how \emph{limited information}, in particular, uncertainty in firms’ beliefs about competitors’ demand intercept $a_j$, alters market outcomes and regulates price coordination. We analyze both settings where firms possess perfect information and where they operate under imperfect information, using either estimators or noisy signals to form beliefs about competitor demand parameters. Throughout, we distinguish clearly between an estimator (a generic mapping from data to belief) and a signal (a particular realization with randomness). Noisy signals can be viewed as a special case of estimators, and will be introduced later.

Using this model as a foundation, we explore a sequence of progressively richer scenarios that build toward our central policy insight:

\textbf{Information Distortion (Section~\ref{sec:distortion})}: We first examine how imperfect information—modeled as biased or noisy estimates of rivals’ baseline demand—distorts firms’ pricing, profit, and consumer welfare.

\textbf{Cartel Dynamics (Section~\ref{sec:cartel})}: Next, we model pricing coordination within a cartel as a social learning process \`a la \citet{degroot}, where firms iteratively update their price proposals toward a shared target. This characterizes how decentralized (tacit) coordination may lead to supracompetitive outcomes.

\textbf{Tradeoff Analysis (Section~\ref{sec:tradeoff})}: Finally, we evaluate the use of controlled information friction as a regulatory lever to fight against (tacit) price coordination. We derive upper bounds on noise levels that limit harm to non-collusive firms and consumers, and quantify how such noise delays cartel consensus. This yields an efficient frontier for information-based intervention.

\section{Baseline Equilibria: How Should Firms Price?} \label{sec:equilibria}

In this section, we characterize equilibrium pricing outcomes under two benchmark information regimes:
(i) one in which firms have perfect information about all market demand parameters, and
(ii) one in which firms operate under imperfect information, relying on fixed estimates of competitors’ demand conditions.

These two regimes give rise to two distinct equilibrium concepts: a Nash equilibrium under full information, and a Bayesian Nash equilibrium\footnote{Bayesian Nash equilibrium refers to games with incomplete information, where players optimize given beliefs. We do not assume Bayesian updating—only fixed estimates.} under partial information. Both will serve as foundational primitives for the modeling extensions that follow.

\subsection{Nash Equilibrium with Perfect Information}

Assume \emph{symmetric} and \emph{complete} information, meaning that all demand information $\inset{a_i,b_i, c}_{i\leq 2}$ are public. Firm $F_i$'s profits are given by
\begin{align}\label{eqn:profit}
    \pi_i\inparen{p_i, p_j} := \inparen{p_i - k_i} \cdot D_i\inparen{p_i, p_j},
\end{align}
profit per unit times the demand. Firm $F_i$ set price $p_i$ to maximize profit. Their optimization can be written as:
\begin{align}
    \max_{p_i \geq 0} \quad \inparen{p_i - k_i} \cdot D_i\inparen{p_i, p_j}.
\end{align}
Without loss of generality, we assume zero production cost for each firm (set $k_i=0$), since cost appears linearly in the profit function and do not meaningfully alter the strategic structure. We have the following result: 
\begin{lemma} \label{lem:opt_p_sym}
When two competing firms face linear demand, and assuming all demand information are common knowledge, each firm’s optimal pricing strategy—anticipating the competitor’s response—is given by:
\begin{align}
    p_i^N = \frac{2 a_i b_j + c a_j}{4 b_i b_j - c^2},\quad
    p_j^N = \frac{2 a_j b_i + c a_i}{4 b_i b_j - c^2}.
\end{align}
These prices constitute a symmetric Nash equilibrium.
\end{lemma}

Intuitively, this equilibrium reflects rational price-setting behavior under complete information: each firm sets its price anticipating the strategic response of the competitor. Since the firms' demand functions are interdependent, each firm's optimal price depends not only on its own demand but also on those of its rival. 

Lemma~\ref{lem:opt_p_sym} shows that the equilibrium balances competitive pressure (captured by cross-price sensitivity $c$) and the firm-specific market conditions (via $\inset{a_i, b_i}_{i\leq 2}$). All proofs are provided in Appendix~\ref{app:proofs}.

\subsection{Bayesian Nash Equilibrium with Imperfect Information} \label{sec:bne}

We now consider the case where firms have incomplete information about their competitor’s demand. Each firm $F_i$ knows its own parameters $\inset{a_i, b_i}$ and observes $b_j$ and $c$, but does not know the rival’s baseline demand $a_j$. Instead, it uses a fixed estimate $\hat{a}_j$ in place of $a_j$ to reason about the competitor’s behavior.

Importantly, we assume firms are strategically symmetric: each anticipates that its competitor will respond optimally using the same best-response rule as in the complete-information case (Lemma~\ref{lem:opt_p_sym}), but evaluated under \emph{imperfect information}.

This departure from perfect information leads to pricing distortions: each firm optimizes against an estimated environment rather than the true one. A consistent Bayesian Nash equilibrium arises:

\begin{lemma} \label{lem:opt_p_asym}
Under imperfect information, where firm $F_i$ uses $\hat{a}_j$ as a fixed estimate of its competitor’s baseline demand, the resulting prices are:
\begin{align}
p_i^{\BNE} = \frac{2 a_i b_j + c \hat{a}_j}{4 b_i b_j - c^2}, \quad
p_j^{\BNE} = \frac{2 a_j b_i + c \hat{a}_i}{4 b_i b_j - c^2}.
\end{align}
These constitute a Bayesian Nash equilibrium under fixed estimates.
\end{lemma}

This equilibrium mirrors the complete-information case (Lemma~\ref{lem:opt_p_sym}) in form, except that each firm now substitutes an estimate $\hat{a}_j$ for its rival’s true baseline demand. Despite its apparent simplicity, this result arises from a nontrivial self-referential structure: each firm's optimal price depends on its belief about the competitor's response, which itself is derived from an estimate of the firm’s own action. 

Together, these two pricing schemes characterize how firms behave when their strategic environment is either fully known or partially estimated.

\subsubsection*{Interior Feasibility}

The closed-form equilibrium prices in Lemma~\ref{lem:opt_p_asym} are interior solutions. 
Mathematically, however, inaccurate or systematically misspecified estimates $\hat a_j$, 
or extreme noise realizations, may drive a firm’s proposed price too high (beyond its choke point) 
or too low (negative). In such cases the literal expressions cease to be economically feasible: 
a negative price implies exit, while an excessive price implies zero demand. 
To capture these feasibility constraints, we introduce the following convention.

\begin{definition}[Interior Feasibility]\label{def:interior}
An equilibrium profile $(p_i^{\BNE},p_j^{\BNE})$ is said to be \emph{interior} if
\begin{align}
    p_i^{\BNE} \geq 0 \quad\text{and}\quad
    D_i(p_i^{\BNE}, p_j) = a_i - b_i p_i^{\BNE} + c p_j > 0,
\end{align}
for each $i\in\inset{1,2}$ and any rival response $p_j$. 
If $p_i^{\BNE} < 0$, the natural outcome is market exit ($p_i=0$). 
If $p_i^{\BNE} > p_i^{\max}(p_j) := \tfrac{a_i + c\,p_j}{b_i}$, firm $F_i$'s choke price, then firm $i$’s demand vanishes ($D_i=0$). 
We refer to such outcomes as \emph{boundary cases}.
\end{definition}

In what follows we focus on interior equilibria. Boundary violations do not affect the algebraic 
form of best responses.
For clarity of exposition, our main results proceed under Definition~\ref{def:interior}, 
with boundary cases treated separately.

In the next sections, we examine how such estimation errors propagate into market distortions and reshape the dynamics of collusion.

\section{Information Distortion} \label{sec:distortion}
We study how imperfect information distorts market outcomes, specifically, equilibrium prices, firm profits, and consumer welfare. Under the imperfect information regime (Section~\ref{sec:bne}), each firm $F_i$ forms a fixed estimate $\hat{a}_j$ of its competitor’s baseline demand $a_j$, which may be biased, noisy, or even systematically misspecified.

This section proceeds in two parts. First, we derive closed-form expressions showing how estimation error in $\hat{a}_j$ propagates into distortions in pricing, profit, and welfare.

Second, we specialize to the case where firms observe noisy signals of the form $\hat{a}_j = a_j + \eta_j$, where $\eta_j$ is a random noise term. This additive structure reflects common information frictions in algorithmic pricing environments and enables concrete analysis under specific noise distributions such as Laplace or Gaussian. For the remainder of this paper, we adopt noisy signals as our working model of imperfect information.

Throughout, we focus on interior solutions (Definition~\ref{def:interior}) with equilibrium prices strictly between zero and the choke price. Corner cases where $p^{\BNE}$ falls outside this range are relegated to Appendix~\ref{app:cornercases}, where we characterize distortions under market exit ($p^{\BNE} < 0$) and zero demands ($p^{\BNE} > p^{\max}$).

\subsection{Price Distortion} \label{sec:price_distortion}
We begin by examining how limited information alters firm pricing. From Section~\ref{sec:equilibria}, the two equilibrium prices are given by:
\begin{align}
    p_i^{N} &= \frac{2 a_i b_j+c a_j}{4 b_i b_j - c^2} \quad \mbox{(Perfect information)} \\
    p_i^{\BNE} &= \frac{2 a_i b_j+c \hat{a}_j}{4 b_i b_j-c^2} \quad \mbox{(Imperfect information)}
\end{align}

We define the price distortion for firm $F_i$ as $\Delta p_i := p_i^{\BNE} - p_i^{N}$, the gap between its price under imperfect and perfect information.

The perfect information equilibrium (Lemma~\ref{lem:opt_p_sym}) provides a natural benchmark, as it defines what each firm would charge if all demand parameters were known. When information is imperfect, however, firms must rely on estimated values, yielding a different pricing outcome (Lemma~\ref{lem:opt_p_asym}). Comparing $p_i^{\BNE}$ to $p_i^{N}$ brings us to the formal result:

\begin{theorem}[Price Distortion] \label{thm:p_dist}
When firm $F_i$ operates under imperfect information, its optimal price deviates from the complete-information benchmark by
\begin{align*}
    \Delta p_i = \frac{c}{4 b_i b_j - c^2} \cdot (\hat{a}_j - a_j).
\end{align*}
$\Delta p_i$ is positive if firm $F_i$ overestimates its rival’s baseline demand (i.e., $\hat{a}_j > a_j$), and negative if it underestimates ($\hat{a}_j < a_j$). Further, the magnitude of distortion scales linearly with the size of the estimation error and is amplified by the degree of competitive coupling (i.e., as cross-price sensitivity $c$ increases).
\end{theorem}

\noindent\textbf{Economic Interpretation:}
Price distortion under imperfect information is shaped by two key factors: the direction of the estimation error (i.e., $\operatorname{sgn}\inparen{\Delta p_i}$) and the degree of strategic coupling (i.e., $c$) between firms.

First, the direction of distortion is entirely determined by how firm $F_i$ perceives its competitor’s baseline demand. If $F_i$ overestimates its rival's baseline demand ($\hat{a}_j > a_j$), it perceives its rival as having stronger market power, anticipating less aggressive pricing from the competitor, and raises its own price. Conversely, underestimating rival ($\hat{a}_j < a_j$) leads firm $F_i$ to expect more intense price competition, prompting it to lower its price to defend market share.

Second, the magnitude of this pricing distortion grows proportionally with the size of the estimation error ($\abs{\hat{a}_j - a_j}$). While this is intuitive, its consequence is amplified by the cross-price sensitivity ($c$). The more strategically entangled the firms are ($c$ higher), the more sensitive the demand of each firm is to other’s pricing. In such settings, even small misjudgments in estimating rival demand can lead to large swings in pricing behavior. Thus, in highly substitutable markets, limited information not only shifts prices in predictable directions, but also triggers magnified competitive response.

In short, pushing too high risks ceding market share, while pushing too low erodes margins—imperfect information tilts this trade-off.

\subsection{Profit Distortion}

Having seen how limited information distorts firm pricing, we now examine its impact on profit $\pi_i$. Because profits depend on both a firm's own price and its competitor's response, profit distortion arises as a second-order effect that react indirectly through prices. 

For both information schemes, the $F_i$'s profits are denoted as:
\begin{align}
    \pi_i^{N} &= \pi_i\inparen{p_i^{N}, p_j^{N}}  \quad \mbox{(Perfect information)} \\
    \pi_i^{\BNE} &= \pi_i\inparen{p_i^{\BNE}, p_j^{\BNE}} \quad \mbox{(Imperfect information)}
\end{align}

We define the profit distortion as $\Delta \pi_i := \pi_i^{\BNE} - \pi_i^{N}$. Applying a first-order Taylor approximation\footnote{This expression comes from a first-order expansion of $\pi_i$ in $(p_i, p_j)$ around $(p_i^{N}, p_j^{N})$. Since $\pi_i$ is quadratic in prices, the approximation is exact up to second-order terms, which are negligible when imperfect-information prices remain near the perfect-information benchmark.} around the perfect-information equilibrium yields the following result:

\begin{theorem}[Profit Distortion] \label{thm:pi_dist}
When firm $F_i$ operates under imperfect information, its profit deviates from the complete-information benchmark by
\begin{align*}
    \Delta \pi_i \approx \frac{c^2 \left(2 a_i b_j + c a_j\right)}{\left(c^2 - 4 b_i b_j\right)^2} \cdot (\hat{a}_i - a_i).
\end{align*}
$\Delta \pi_i$ is positive if the competitor $F_j$ overestimates firm $F_i$'s baseline demand ($\hat{a}_i > a_i$), and negative if it underestimates. The magnitude scales linearly with the size of rival's estimation error, and is amplified by the competitive coupling (i.e., as $c$ increases). 
Notably, The distortion is independent of $F_i$’s own estimation error.
\end{theorem}

\noindent\textbf{Economic Interpretation:}
Profit distortion under imperfect information reflects how a firm is perceived by its competitor. If rival firm $F_j$ overestimates $F_i$'s baseline demand ($\hat{a}_i > a_i$), believing $F_i$ has far greater market power, it anticipates less aggressive pricing from $F_i$ and softens its own price response, allowing $F_i$ to earn higher profit. Conversely, underestimation leads to aggressive competition and reduced profit for $F_i$.

The magnitude of this effect grows linearly with the size of the rival’s estimation error. It is further amplified by cross-price sensitivity ($c$): in markets where firms’ demands are more entangled, misperceptions by rival firm induce stronger price reactions, with larger profit consequences.

Importantly, profit distortion depends only on how $F_i$ is perceived by its rival $F_j$. At first order, a firm’s own price is a best response to its belief, so only the competitor’s pricing distortion—affecting demand faced by $F_i$—drives profit distortion.

\subsection{Consumer Welfare Distortion}

We now turn to consumer welfare, defined as the surplus enjoyed by buyers under firm $F_i$'s pricing. In linear demand settings, this corresponds to the area under the inverse demand curve and above the posted price. 

\begin{figure}[H]
\centering
\includegraphics[width=0.4\textwidth]{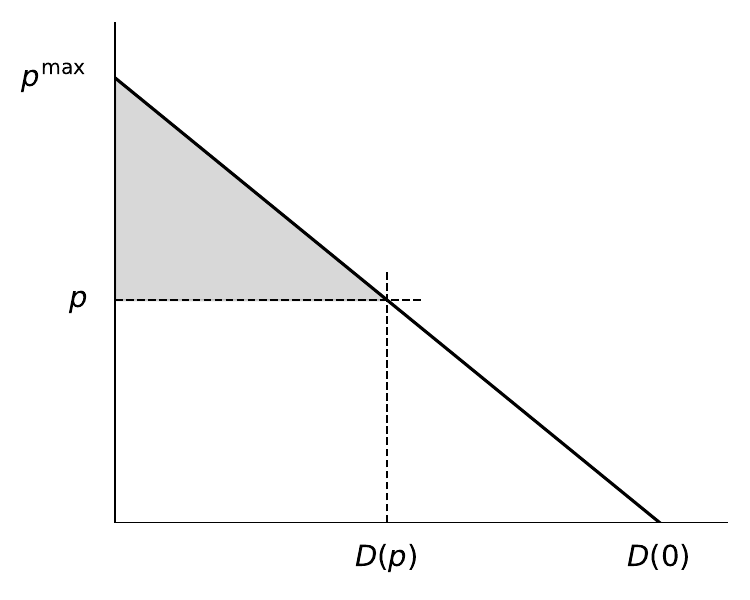}
\caption{Consumer surplus (shaded region) under inverse demand curve.}
\label{fig:consumer_surplus}
\end{figure}

Figure~\ref{fig:consumer_surplus} illustrates firm $F_i$'s consumer surplus, assuming the competitor’s price is fixed to isolate the effect.

The horizontal axis represents demand, and the vertical axis is price. The shaded triangle captures the area between the choke price $p^{\max}$, the price at which demand falls to zero, and the market price $p$, reflecting consumer welfare from purchasing firm $F_i$'s product. The closed-form expression appears below:

\begin{definition}[Consumer Welfare]
The consumer surplus generated by firm $F_i$ is given by
\begin{align}
    U_i(p_i, p_j) = \frac{D_i(p_i, p_j)^2}{2b_i} = \frac{(a_i - b_i p_i + c p_j)^2}{2b_i}.
\end{align}
This corresponds to the area under the inverse demand curve above the market price $p_i$.\footnote{See Appendix~\ref{app:consumer_welfare} for full derivation.}
\end{definition}

Consumers of firm $F_i$ are not a fixed group: at low prices the market includes lower-valuation buyers, while at higher prices only higher-valuation buyers remain. Accordingly, $U_i$ measures the surplus of whichever consumers purchase at the prevailing price.\footnote{Notice that under linear demand with cross-price effects, a high price profile $(p_i,p_j)$ can yield the same demand as a much lower profile; what differs is the cutoff of valuations, i.e., the composition of buyers.}

We define consumer welfare distortion as the difference between the welfare under imperfect and perfect information:
\begin{align*}
    \Delta U_i := U_i^{\BNE} - U_i^N = U_i\inparen{p_i^{\BNE}, p_j^{\BNE}} - U_i\inparen{p_i^N, p_j^N}
\end{align*}
As in the profit case, a first-order Taylor expansion around the perfect-information equilibrium provides tractable insight into how estimation errors impact consumer outcomes:

\begin{theorem}[Welfare Distortion] \label{thm:u_dist}
When firm $F_i$ operates under imperfect information, its consumer welfare increases from the perfect-information benchmark by
\begin{align*}
    \Delta U_i \approx \frac{c \left(2 a_i b_j + c a_j\right)}{\left(c^2 - 4 b_i b_j\right)^2} \cdot \inbrak{c (\hat{a}_i - a_i) - b_i (\hat{a}_j - a_j)}.
\end{align*}

Consumer welfare increases when:
rival firm $F_j$ overestimates $F_i$’s baseline demand ($\hat{a}_i > a_i$), or
when firm $F_i$ underestimates its rival's baseline demand ($\hat{a}_j < a_j$).

The effects reverse when the inequalities are flipped. Further, the distortion effect is amplified as cross-price sensitivity $c$ increases. 
\end{theorem}

\noindent\textbf{Economic Interpretation:}
Unlike profit distortion, which arises solely from how a firm is perceived by its rival (i.e., $\hat{a}_i - a_i$), consumer welfare distortion is influenced by both a firm’s own estimation of its competitor and how the competitor perceives it.

Specifically, the term $(\hat{a}_i - a_i)$ captures how the rival’s belief affects their price $p_j$, which shifts $F_i$’s demand curve and therefore the area under it. The term $(\hat{a}_j - a_j)$ reflects how $F_i$'s mis-estimation of rival alters its own price. Both enter at the first order, unlike profit distortions which depend only on rival perception. This dual exposure makes consumer surplus strictly more fragile to informational frictions than firm profit. 

The implications are twofold. In markets with strong substitutability ($c$ large), both channels are magnified, so small errors scale into large welfare swings. And, perhaps surprisingly, noise can act as a pro-competitive force: prices may fall below benchmark levels, boosting consumer surplus. Yet distortions can also cut the other way. 

Finally, we note that these results fall outside the interior benchmark in extreme corner cases: as shown in Appendix~\ref{app:cornercases}, market exit ($p_i^{\BNE} = 0$) may raise or lower welfare depending on the rival’s response, whereas zero demand ($p_i^{\BNE} > p_i^{\max}$) beyond the choke price yields an unambiguous loss.

\subsection{Noisy Signal as Estimator}
Thus far, we have examined how limited information affects firm behavior: altering prices at first order, and distorting profits and consumer welfare at second order.

We now specialize to a fixed form of imperfect information: noisy signals, where each firm observes a noisy signal of its rival's baseline demand. As prefaced in Section~\ref{sec:linear_demand} (General Model), noisy signals are specific realizations of estimators and serve as a natural bridge between theory and real-world informational frictions. From this point forward, we focus exclusively on this signal-based framework for its analytical clarity and empirical relevance.

\textbf{Modifications to base expressions (interior cases):} Formally, suppose firm $F_i$ observes rival $F_j$'s baseline demand signals of the form $\hat{a}_j =  a_j + \eta_j$, where $\eta_j \overset{iid}{\sim} F$, some distribution, as estimation. Substituting the notation of estimation error $(\hat{a} - a)$ with $\eta$, then the distortions are given by:
\begin{align}
    \Delta p_i &= \frac{c}{4 b_i b_j-c^2} \cdot \eta_j \quad \mbox{(Price)} \label{eqn:p_dist_sig}\\
    \Delta \pi_i &\approx  \frac{c^2 \left(2 a_i b_j+c a_j\right)}{\left(c^2-4 b_i b_j\right){}^2} \cdot \eta_i \quad \mbox{(Profit)} \\
    \Delta U_i &\approx \frac{c \left(2 a_i b_j+c a_j\right)}{\left(c^2-4 b_i b_j\right){}^2} \cdot \left( c \cdot \eta_i -b_i \cdot \eta_j\right) \quad \mbox{(Welfare)} \label{eqn:u_dist_sig}
\end{align}

These formulas characterize realized-signal distortions: price distortion is exact, while the profit and welfare distortions follow from the Taylor expansion (Theorems~\ref{thm:pi_dist}~and~\ref{thm:u_dist}).

We begin by analyzing the effect of realized noise on pricing, welfare, and profit, up to first-order precision. We then simulate full distributions of noise to assess the magnitude and risk of distortion in exact form.

As established in our baseline equilibria, infeasible prices leads to corner cases of market exit and zero demand. Our realized-signal analysis assumes interior prices (Definition~\ref{def:interior}) and thus abstracts from truncation, whereas our distributional simulations impose it since they consider the full support of signal realizations.

\subsubsection*{How Realized Noises Affect Outcomes}
We summarize how realized noisy signal in baseline demand estimates affects firm $F_i$'s equilibrium outcomes. These results follow immediately from Theorems~\ref{thm:p_dist}, \ref{thm:pi_dist}, and \ref{thm:u_dist} under the noisy signal notation. Table~\ref{tab:eta_effects} reports the direction of distortion in price, consumer welfare, and firm profit as each noise term $\inparen{\eta_i, \eta_j}$ increases or decreases.

\begin{table}[H]
\centering
\begin{tabular}{lcccc}
\toprule
\textbf{Distortion Type} 
& $\eta_i \nearrow$ 
& $\eta_i \searrow$ 
& $\eta_j \nearrow$ 
& $\eta_j \searrow$ \\
\midrule
$\Delta p_i$ (Price)     
& NA                   
& NA                  
& $\uparrow$        
& $\downarrow$      \\
$\Delta \pi_i$ (Profit)   
& $\uparrow$        
& $\downarrow$      
& NA                 
& NA                  \\
$\Delta U_i$ (Welfare)   
& $\uparrow$        
& $\downarrow$      
& $\downarrow$      
& $\uparrow$        \\
\bottomrule
\end{tabular}
\captionsetup{position=bottom}
\caption{Impact of signal noise on firm $F_i$'s price, welfare, and profit distortions. 
$\eta_i$ and $\eta_j$ are noise terms in the signals used by firms to estimate each other's demand. Outcomes reflect distortions from $F_i$'s perspective only.}
\label{tab:eta_effects}
\end{table}

Price and profit distortions load on different noise terms. Firm $F_i$’s price distortion $\Delta p_i$ depends on $\eta_j$ because its best response is shaped by how it perceives rival $F_j$’s demand; errors in $a_j$ shift $p_i$ directly. Profit distortion $\Delta \pi_i$, on the other hand, depends on $\eta_i$, since realized profit is determined by how accurately rival $F_j$ estimates $F_i$’s own demand intercept $a_i$: misestimation shifts $p_j$, altering $F_i$’s residual demand and thus its profits.

A second observation is that profit and welfare always move in the same direction with $\eta_i$. When firm $F_j$ overestimates $F_i$’s baseline demand $a_i$, it sets a higher price $p_j$, which relaxes competition. This expands $F_i$’s realized demand $D_i$, boosting both its profit $\pi_i$ and the surplus enjoyed by its customers. Conversely, when $F_j$ underestimates $a_i$, it prices more aggressively, reducing $D_i$ and driving down both profit and welfare. The alignment arises because both profit and welfare depend directly on $F_i$’s realized demand, which in turn is shaped by how $F_j$’s estimate $a_i+\eta_i$ affects its pricing choice.

\subsubsection*{Random Noise, Real Risk}

To this point, we have examined how realized distortions respond to specific values of signal noise. A natural next step is to simulate what happens when noise itself is drawn from a full distribution.

The harm of information distortion arises not merely from its realized magnitude, but from its fluctuations that can drive realized outcomes into regions of significant welfare or profit loss, even if the average effect vanishes under mean-zero noise. Large deviations in either direction can also drive the market collapse (see Appendix~\ref{app:cornercases}). 

Figure~\ref{fig:distortion_noise} illustrates this dynamic in detail. We simulate two mean-zero noise schemes, Laplace and Gaussian, each parameterized by a common scale parameter $\sigma$. For Gaussian noise, the variance is $\sigma^2$; for Laplace noise, the variance is $2\sigma^2$. Unlike the Taylor-based approximations earlier, these simulated distortions are computed exactly from the model (with truncation applied).\footnote{We impose minimal feasibility only:
(i) a price floor $\max\inset{0,p_i^{\BNE}}$; no upper cap on $p_i$ is needed;
(ii) a demand floor $\max\inset{0,D_i(p_i^{\BNE},p_j^{\BNE})}$ so that prices above the choke level imply zero trade.
Binding cases map to the corner analyses in Appendix~\ref{app:cornercases} (market exit at $p_i=0$; zero demand when $p_i$ exceeds the choke price).}

\begin{figure}[H]
\centering
\begin{subfigure}{0.48\textwidth}
    \centering
    \includegraphics[width=\linewidth]{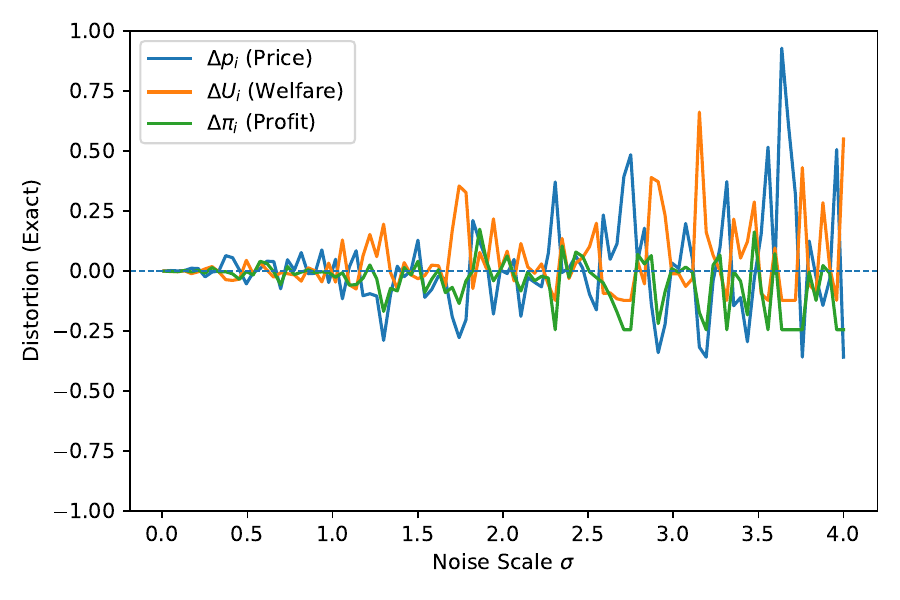}
    \caption{Laplace Noise}
\end{subfigure}
\hfill
\begin{subfigure}{0.48\textwidth}
    \centering
    \includegraphics[width=\linewidth]{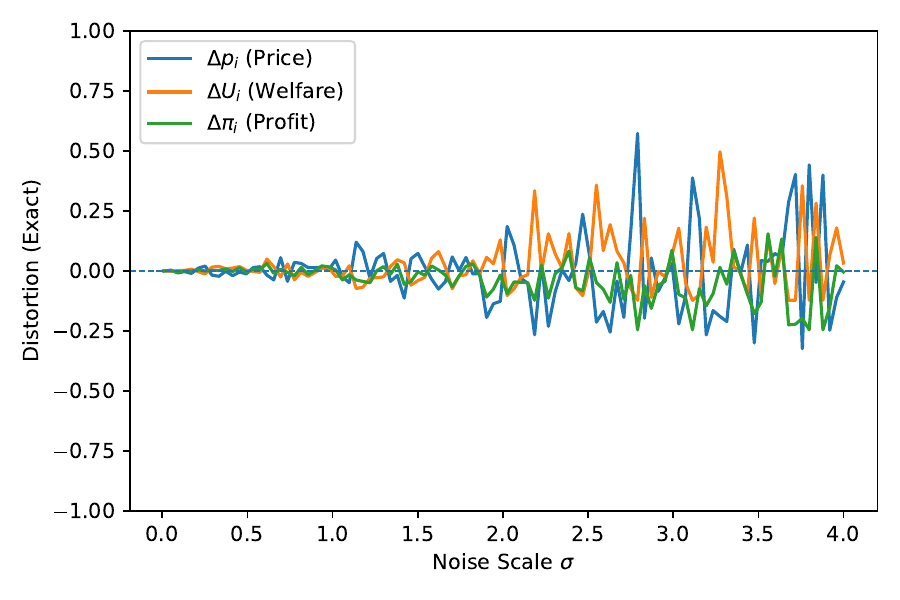}
    \caption{Gaussian Noise}
\end{subfigure}
\caption{Exact distortions in price ($\Delta p_i$), consumer welfare ($\Delta U_i$), and profit ($\Delta \pi_i$) under different noise types. Each point represents a single realization of the noise term at scale $\sigma$.}
\label{fig:distortion_noise}
\end{figure}

As $\sigma$ increases, the variance of the noise increases. This rise in variance leads to greater fluctuation in realized distortions. While both noise types exhibit the same qualitative pattern of increasing deviation, Laplace noise tends to produce more extreme absolute distortions due to its heavier tails.

\begin{figure}[H]
\centering
\begin{subfigure}{0.48\textwidth}
    \centering
    \includegraphics[width=\linewidth]{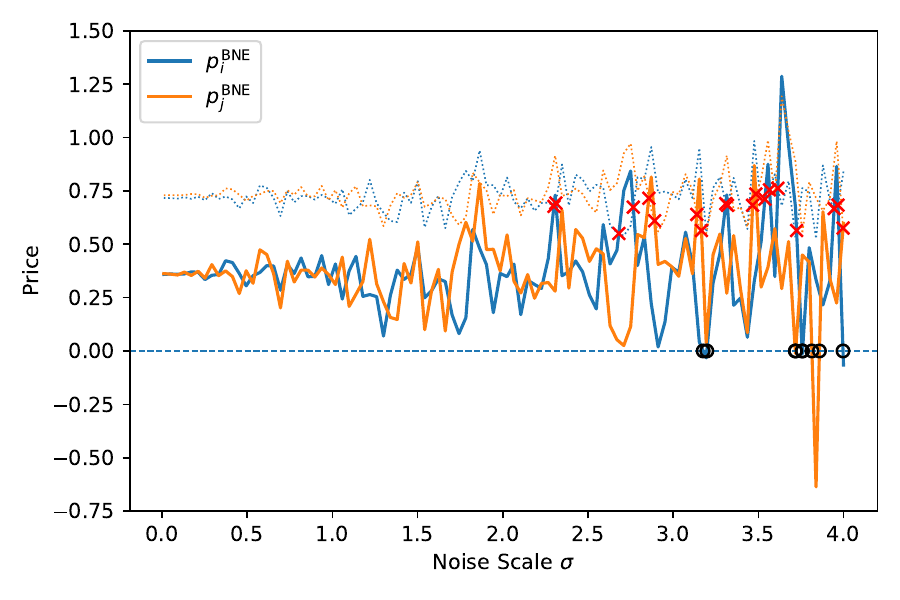}
    \caption{Laplace Noise}
\end{subfigure}
\hfill
\begin{subfigure}{0.48\textwidth}
    \centering
    \includegraphics[width=\linewidth]{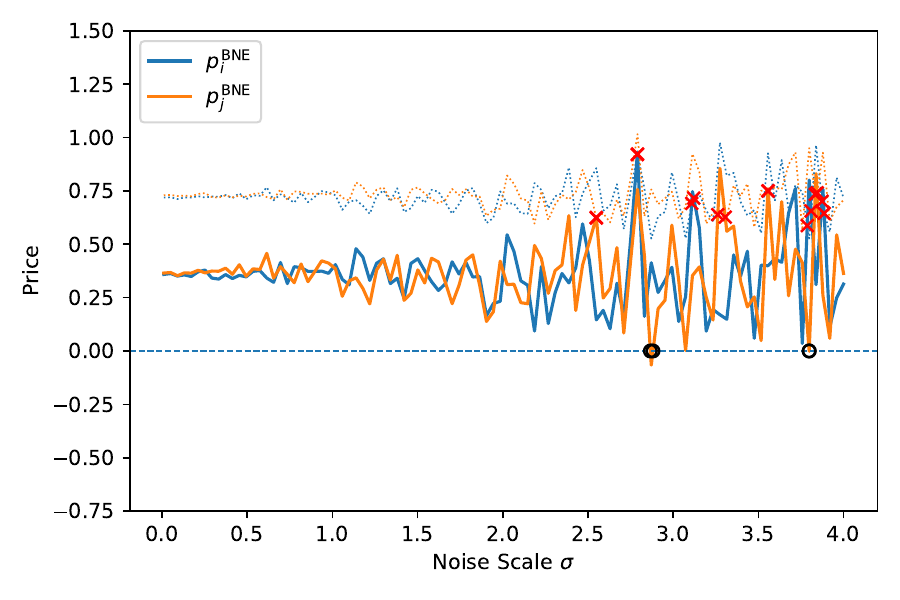}
    \caption{Gaussian Noise}
\end{subfigure}
\caption{Posted prices $p_i^{\text{BNE}}$ and $p_j^{\text{BNE}}$ (untruncated) under different noise types. 
Each point represents a single realization at scale $\sigma$. Solid curves are $p_i^{\text{BNE}}$ and $p_j^{\text{BNE}}$; 
dotted lines show the corresponding choke prices $p_i^{\max}$ and $p_j^{\max}$. 
Crosses denote choke crossings, where demand collapses to zero. 
Hollow circles denote zero crossings, where negative prices imply effective market exit (the firm must pay consumers).}
\label{fig:price_paths}

\end{figure}

Beyond distortions, it is also useful to examine posted prices under noise ($p^{\BNE}$). These may occasionally fall below zero or exceed the choke price, violating interiority (Definition~\ref{def:interior}). Figure~\ref{fig:price_paths} illustrates this using the same setup as in Figure~\ref{fig:distortion_noise}. We simulate untruncated firm prices $p_i^{\text{BNE}}$ and $p_j^{\text{BNE}}$ (solid lines) under Laplace and Gaussian noise as the noise scale $\sigma$ increases, together with their corresponding choke prices (dotted lines) in the same colors.

Under Laplace noise, heavier tails push prices to thread both thresholds at high $\sigma$. Under Gaussian noise, price movements remain relatively more contained, with fewer boundary crossings observed. In sum, large fluctuations can amplify the risk of significant welfare or profit loss, and, in extreme cases, even drive the market to collapse.

\section{Cartel Dynamics}
\label{sec:cartel}

In this section, we lay out the coordination mechanism behind tacit collusion. While previous sections examine how limited information distorts competitive outcomes, we now study how a cartel may reach consensus on collusive pricing under imperfect information.

Our analysis builds on models of social learning and belief updating, specifically, the DeGroot process \citep{degroot, degroot1973}, which models opinion averaging. Though originally designed to describe group belief formation, its extensions have been used to study decentralized coordination under influence and misinformation \citep{acemoglu2009opinion}. 

In our context, this offers a tractable abstraction of tacit algorithmic collusion: firms update prices iteratively without direct communication, eventually converging to a common (supracompetitive) target.

\subsection{Belief Averaging in a Duopoly}
\label{sec:degroot_crowd}

We begin with a simple belief averaging process. Consider two firms, $F_i$ and $F_j$, that repeatedly update their proposed prices by averaging their current beliefs. Let $x_i(t)$ denote the proposed price of firm $F_i$ at time $t$. At each round, firm $F_i$ places weight $\alpha \in (0,1)$ on its own previous belief and $1 - \alpha$ on its rival’s:
\begin{align}
    x_i(t+1) = \alpha x_i(t) + (1-\alpha)x_j(t)
    \quad \Rightarrow \quad
    x(t+1) = A x(t), \quad A = 
    \begin{bmatrix}
        \alpha & 1-\alpha \\ 
        1-\beta & \beta
    \end{bmatrix}
\end{align}
The sequence $x(t)$ converges to a shared consensus (i.e., both firms eventually propose the same price) under mild conditions.\footnote{Weight matrix $A$ is aperiodic and strongly connected.} This consensus reflects a stable collusive outcome reached through mutual adaptation.

In cartel settings, however, consensus may be shaped by asymmetric influence. For example, firms pushes for higher prices may be more persistent or persuasive, leads to leader-follower dynamics. This motivates the \emph{forceful agent} modification of \citet{acemoglu2009opinion}, where one firm is unyielding and never updates (e.g., leader):
\begin{align}
    \forall t,\; x_j\inparen{t+1}=x_j\inparen{t} \implies D = \begin{bmatrix}
        \alpha & 1-\alpha \\ 
        0 & 1
    \end{bmatrix}
\end{align}
Here, firm $F_j$ rigidly holds its position, while $F_i$ adjusts toward it over time. This abstraction captures the idea that cartel negotiations often gravitate toward the most aggressive member, typically the firm pushing for the highest price.

\subsection{Cartel Consensus with Noisy Inputs}
\label{sec:cartel_consensus}

We now formalize this stylized cartel pricing dynamic. Cartel members are \emph{selfish} and \emph{trustless}, self-maximizing and unwilling to share their private demand conditions. Instead of immediately adopting a collusive outcome, they engage in an implicit (tacit) bargaining process: adjusting their price proposals over time in response to public signals until consensus is reached.

Each firm knows its own baseline demand $a_i$, but not others’. Prices $p_i$ and parameters $b_i$, $c$ are public. This information setting mirrors the incomplete-information scheme (see Section~\ref{sec:bne}). 

We model the cartel negotiation as a three-phase process:

\begin{enumerate}[1.]
   \item \textbf{Signal Observation.} Each firm’s true baseline demand $a_i$ is private and observed imperfectly as a noisy public signal: $s_i = a_i + \eta_i$, where $\eta_i \overset{iid}{\sim} F$ for some distribution $F$.\footnote{One interpretation is to view this as a shared pricing algorithm $\mathcal{A}$ that aggregates pooled market data and releases sanitized demand estimates. Rather than granting firms direct access to rivals’ conditions, $\mathcal{A}$ provides a noisy signal, consistent with how cartels might prefer a ``trustless” intermediary to coordinate expectations.}

    \item \textbf{Initial Pricing and Target Formation.} Each firm computes its own equilibrium price $p_i^{\BNE}$ using its private $a_i$ and the noisy signals $s_j$ from its rival. The cartel selects a supracompetitive target price\footnote{One could also imagine collusive prices set strictly above 
$\max\{p_i^{\BNE},p_j^{\BNE}\}$. This would not meaningfully alter our subsequent analyses, but adopting $\max$ ensures tractability by pinning 
down a unique collusive outcome.} as the maximum across firms:
    \[
        p^{\text{target}} := \max\inset{p_i^{\BNE}, p_j^{\BNE}}.
    \]
    
    \item \textbf{Deterministic Update.} Each firm updates its proposed price by gradually adjusting toward $p^{\text{target}}$ at rate $\alpha$:
    \[
        x_i(t+1) = (1-\alpha)\cdot x_i(t) + \alpha \cdot p^{\text{target}}.
    \]
    Consensus is reached when $\max_{i \le 2} |x_i(t) - p^{\text{target}}| < \epsilon$, for some tolerance $\epsilon>0$.
\end{enumerate}
 This setup captures a social learning dynamic with forceful agent. Table~\ref{tab:noisy_model_timeline} summarizes the timestep:

\begin{table}[H]
\centering
\begin{tabular}{@{}llp{9cm}@{}}
\toprule
\textbf{Time} $t$ & \textbf{Phase} & \textbf{Description} \\
\midrule
1 & Initialization &
Firm $F_i$ obtains private $a_i$. \\
\midrule
2 & Signal Sharing \& Nash Pricing &
$F_i$ receive $s_j = a_j + \eta_j$, then compute $p_i^{\BNE}$ using rival’s noisy signal. \\
\addlinespace
& Target Formation &
Cartel sets $p^{\text{target}} = \max_i \{p_i^{\BNE}\}$. \\
\midrule
3+ & Updates Rounds &
Each firm starts from $x_i(0) = P_i^N$, then updates toward $p^{\text{target}}$. \\
\addlinespace
& Consensus Check &
Consensus if $\max_i |x_i(t) - p^{\text{target}}| < \epsilon$. \\
\bottomrule
\end{tabular}
\caption{Timeline of the Cartel Consensus Model}
\label{tab:noisy_model_timeline}
\end{table}

\section{Tradeoff Analysis}
\label{sec:tradeoff}
In this section, we ask a regulatory question: can information frictions be used intentionally to deter algorithmic collusion?

We consider a concrete intervention: a regulator requires controlled noise to be injected into the pooled raw market data that feeds pricing algorithms. This stylized setup captures an emerging policy lever: regulating information quality to shape strategic behavior among algorithmic firms.

Formally, we model this as each firm observing a noisy signal of its rival's baseline demand:
\begin{align}
\hat{a}_j = a_j + \eta_j, \quad \eta_j \sim \operatorname{Laplace}(\sigma),
\end{align}
where the noise scale $\sigma$ governs information fidelity, with $\operatorname{Var}(\eta_j) = 2\sigma^2$. This may arise, for instance, when a platform-level pricing algorithm $\mathcal{A}$ aggregates firm demand data $(a)$ and releases a sanitized version $(\hat{a})$ to reduce manipulation or preserve privacy.

Laplacian noise is a natural choice: it is widely adopted in algorithmic systems for balancing fidelity and obfuscation, and has provable optimality in differential privacy settings \citep{laplace}. While our context is not privacy per se, we similarly seek to obfuscate precise demand signals in order to prevent precise coordination.

We structure the policy analysis in three parts. First, we quantify the incentives and consequences of collusion. Second, we formally prove that increasing noise slows down cartel convergence by disrupting the consensus process. Third, we derive upper bounds on noise levels that ensure limited harm to competitive firms and consumer welfare. Together, these results characterize a feasible zone of regulatory intervention: enough noise to deter collusion, but not so much as to destabilize markets.

\subsection{Collusive Incentives and Consequences}

Naturally a first-order question arises: why collusion is attractive to firms and potentially harmful to consumers? The following results characterize both the profit incentive and consumer impact of a cartel adopting a supracompetitive price.

For simplicity, we work with generic price profiles $(p_i,p_j)$ rather than explicit Bayesian Nash notations, adopting a reduced-form perspective on cartel outcomes. We maintain \emph{interior feasibility} (Definition~\ref{def:interior}) throughout, since it is economically implausible for a cartel to coordinate on negative prices or knowingly set prices that induce negative demand.

Cartel consensus is typically shaped by a leader, 
reflecting the self-maximizing nature of firms that push toward their preferred price. 
To capture whether such consensus remains feasible for all firms, we introduce:

\begin{definition}[Consensus-Interior]\label{def:consensus_interior}
Given an initial interior profile $(p_i,p_j)$, we say consensus play is \emph{consensus-interior} if the cartel’s target price $p^{\text{target}}=\max\{p_i,p_j\}$ also leaves both markets interior:
\[
D_k(p^{\text{target}},p^{\text{target}}) > 0 \quad \text{for each } k\in\{i,j\}.
\]
\end{definition}
Unless otherwise noted, all results in this section are stated under
Definition~\ref{def:consensus_interior}.
If this condition fails, consensus drives the lower-priced firm’s demand to zero, a boundary case that yields trivial profits and strict welfare losses for that firm’s consumers (see Appendix~\ref{app:cornercases}).

\begin{theorem}[Incentive to Collude] \label{thm:collusive_incentive}
Suppose both firms adopt a common (supracompetitive) price $p^{\text{target}} = \max\{p_i, p_j\}$, where $p_i$ and $p_j$ are the firms’ initial equilibrium prices under imperfect information. Then, the firm that proposed the higher price ($F_k$) strictly benefits from coordination:
\begin{align}
    \Delta \pi_k^{\text{coll}} := \pi_k(p^{\text{target}}, p^{\text{target}}) - \pi_k(p_k, p_{-k}) > 0
\end{align}
for some $k \in \{i, j\}$.
\end{theorem}

The firm with the higher initial price (e.g., the cartel leader) strictly benefits from consensus at the supracompetitive level. 
For the lower-priced firm (e.g., the follower), its gain is ambiguous if consensus remains interior, but if consensus pushes its demand to zero, its profit collapses to zero. In all cases, at least one firm has a strict incentive to coordinate. As a result, firms are motivated not only to propose higher prices, but also to manipulate rivals into accepting or converging toward those higher prices. Such manipulative dynamics are not speculative; they are central concerns in recent work on algorithmic 
collusion and auditability \citep{hartline2025algoregrefined}.

\begin{theorem}[Consumer Harm under Collusion] \label{thm:collusive_harm}
Suppose both firms adopt a common (supracompetitive) price 
$p^{\text{target}}=\max\{p_i,p_j\}$, where $p_i$ and $p_j$ are the firms’ initial 
equilibrium prices under imperfect information. Then the firm that initially posted 
the lower price ($F_{-k}$) experiences a strict decline in consumer welfare:
\begin{align}
    \Delta U_{-k}^{\mathrm{coll}}
:= U_{-k}(p^{\text{target}},p^{\text{target}})
   - U_{-k}(p_{-k},p_k) \;<\; 0,
\end{align}
while the higher-price firm’s consumers strictly gain.
\end{theorem}

Collusion redistributes consumer welfare asymmetrically. Under our baseline coordination rule
$p^{\text{target}}=\max\{p_i,p_j\}$, the leader’s consumers need not be worse off. 
A natural question arises: if the leader’s own price does not fall (and may even rise under other coordination), why would its consumers benefit? The reason is that the composition of consumers is dynamic: as the rival’s price increases, more buyers shift toward the leader, so the set it serves tilts toward higher-valuation consumers. This resembles the shift from an everyday OTC product to a luxury good—while the good itself becomes costlier in relative terms, the clientele that remains or enters values it more, and measured consumer surplus rises. By contrast, the follower’s consumers categorically lose: their price is raised or they hit the choke price, shrinking their surplus regardless of how the clientele reshuffles (cf. Appendix~\ref{app:cornercases}).\footnote{If the cartel instead sets $p^{\text{target}}>\max\{p_i,p_j\}$, the leader’s consumers face offsetting forces: a rival’s price increase (which boosts their demand) and an own-price increase (which reduces it). The net effect depends on the relative magnitudes of $b$ and $c$. This does not affect our qualitative conclusion that the follower’s consumers strictly lose, so some consumers are necessarily harmed under collusion.}

These results clarify the first-order regulatory concern: firms have clear incentive to collude, and consumers are demonstrably worse off as a result.

\subsection{Noise Slows Down Cartel Convergence}

We show that zero-mean Laplacian noise, used as a policy lever, can significantly delay cartel convergence to a supracompetitive price.

For clarity, let's review the cartel dynamics from Section~\ref{sec:cartel}. First, $p_i^{\BNE}$, the posted price of firm $F_i$ under imperfect information (Lemma~\ref{lem:opt_p_asym}), is computed based on a noisy signal $s_j = a_j + \eta_j$ of its rival’s demand intercept, where $\eta_j \overset{iid}{\sim} \operatorname{Laplace}(\sigma)$. Here, $\eta_i$ and $\eta_j$ are interpreted as a tunable policy lever that reduces firms’ signal precision, then 
\begin{align}
    p_i^{\BNE} &= \frac{2 a_i b_j + c s_j}{4 b_i b_j - c^2}
    = \underbrace{\frac{2 a_i b_j + c a_j}{4 b_i b_j - c^2}}_{=: A_i} + \underbrace{\frac{c}{4 b_i b_j - c^2}}_{=: K} \cdot \eta_j
    = A_i + K \eta_j
\end{align}

The cartel then selects a target price $p^{\text{target}} = \max\inset{p_i^{\BNE}, p_j^{\BNE}}$ and initiates price coordination with $x_i(0) = p_i^{\BNE}$ via:
\begin{align}
    x_i(t+1) = (1 - \alpha) x_i(t) + \alpha p^{\text{target}} 
    &\implies x_i(t) = (1 - \alpha)^t p_i^{\BNE} + \left( 1 - (1 - \alpha)^t \right) p^{\text{target}}
\end{align}
The deviation from consensus evolves as:
\begin{align}
    x_i(t) - p^{\text{target}} = (1 - \alpha)^t (p_i^{\BNE} - p^{\text{target}})
\end{align}

\begin{definition}[$\epsilon$-Consensus]
We say the cartel reaches $\epsilon$-consensus at round $t$ if $\abs{x_i(t) - p^{\text{target}}} < \epsilon$ for all $i$.
\end{definition}

We now derive a probabilistic bound on the deviation from the target price. This quantifies how noise in initial price estimates slows convergence.

\begin{lemma}[Deviation Bound under Laplacian Noise] \label{lem:deviation_bound}
Let $\eta_i, \eta_j \overset{iid}{\sim} \operatorname{Laplace}(\sigma)$ be zero-mean noise in firms’ signal estimates. Then for any tolerance $\epsilon > 0$ and time $t \geq 0$, the deviation from consensus satisfies:
\begin{align}
    \p\left(\abs{x_i(t) - p^{\text{target}}} \geq \epsilon \right)
    \leq \exp\left( - \frac{ \epsilon }{ (1 - \alpha)^t K \sigma } \right),
\end{align}
where $K = \frac{c}{4 b_i b_j - c^2}$.
\end{lemma}

The bound shows that consensus becomes more likely over time: with more negotiation rounds $t$, the probability of firms drifting far from the agreed target shrinks exponentially. A larger tolerance $\epsilon$ also makes consensus more likely, since small disagreements are ignored. By contrast, stronger noise $\sigma$ makes consensus less reliable, raising the chance that firms stray away from the target.

While Lemma~\ref{lem:deviation_bound} characterizes how likely the cartel fails to achieve $\epsilon$-consensus at each fixed time $t$, regulators may instead wish to guarantee convergence within a prescribed error $\epsilon$ with high confidence. This leads us to invert the deviation bound and derive a delay threshold: the minimum number of rounds $t$ needed to ensure $\epsilon$-consensus occurs with probability at least $1 - \delta$ for some $\delta \in (0,1)$.

\begin{theorem}[High-Probability Delay from Laplacian Noise] \label{thm:noise_delay}
Fix tolerance $\epsilon > 0$ for $\epsilon$-consensus, confidence level $1 - \delta \in (0,1)$, and update rate $\alpha \in (0,1)$. Under Laplacian noise of scale $\sigma$, $\epsilon$-consensus is reached with probability at least $1 - \delta$ only after
\begin{align}
    t \geq \log_{1 - \alpha} \left( \frac{ \epsilon }{K \sigma (-\log \delta) } \right),
\quad \text{where } K = \frac{c}{4b_i b_j - c^2}
\end{align}
Larger noise levels $\sigma$ increase this lower bound, implying slower cartel convergence with high probability.
\end{theorem}

\begin{figure}[H]
\centering
\includegraphics[width=0.55\textwidth]{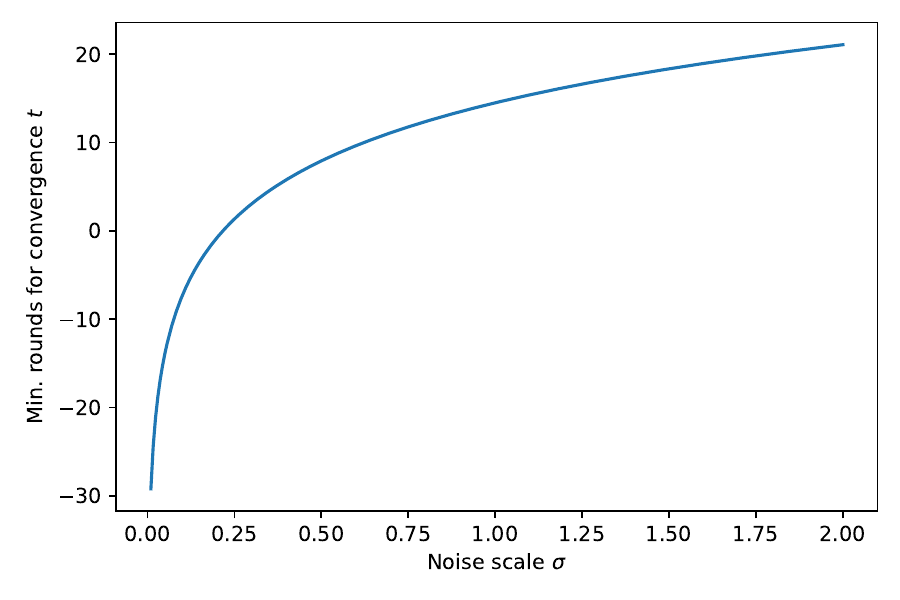}
\caption{Minimum rounds $t$ required to reach $\epsilon$-consensus with high probability ($1 - \delta$), as a function of noise scale $\sigma$.}
\label{fig:laplace_delay}
\end{figure}

Figure~\ref{fig:laplace_delay} illustrates the bound in Theorem~\ref{thm:noise_delay} across different noise levels $\sigma$. The delay to $\epsilon$-consensus grows quickly as informational noise increases.

This result amounts to a central policy insight: introducing carefully designed noise into algorithmic pricing systems can delay supracompetitive coordination. Unlike traditional antitrust tools that intervene ex post \citep{hartline2024algoreg, hartline2025algoregrefined, baranek2025detection, massarotto2025detecting}, this friction operates ex ante and requires no detection of intent or outcome. Notably, the delay is backed by a theoretical guarantee (Theorem~\ref{thm:noise_delay}), making it a robust and practical mechanism for slowing down cartel formation in algorithmic markets.

\subsubsection*{Robustness to Repeated Queries}
So far, we assumed each firm observes a single noisy demand signal. However, in more permissive environments, arguably the worst-case from a regulatory perspective, firms may obtain multiple observations (e.g., repeated queries to a platform-level pricing algorithm) to refine their estimates. 

Concretely, suppose firms query a regulatory interface or pricing algorithm $\mathcal{A}$ multiple times to receive $n$ i.i.d.\ signals of the form $s_j^{(k)} = a_j + \eta_j^{(k)}$.

Let $\bar{s}_j = \frac{1}{n} \sum_{k=1}^n s_j^{(k)}$ denote the sample mean estimator constructed by the cartel to improve information quality. While this estimator is weakly consistent (due to mean-zero Laplacian noise) and asymptotically normal, it retains nontrivial noise for any finite $n$-observations:
\begin{align}
    \bar{s}_j \approx \mathcal{N}\inparen{a_j, \frac{2\sigma^2}{n}}\;\text{for large $n$}
\end{align}

We emphasize that the sample mean is among the simplest possible, yet effective, estimators. Our results thus provide a conservative benchmark:

\begin{proposition}[Persistence of Noise under Repeated Queries] \label{prop:repeat_queries}
Even if firms observe $n$ i.i.d.\ noisy signals from repeated queries to the pricing interface $\mathcal{A}$, the resulting sample mean estimator $\bar{s}_j$ retains residual noise. As a result, cartel convergence remains delayed in finite time.
\end{proposition}

\subsection{Bounding Harm to Market Participants}

Regulatory noise can indeed slow down cartel coordination, even under repeated queries, making controlled information quality an effective policy lever (Theorem~\ref{thm:noise_delay}, Proposition~\ref{prop:repeat_queries}). But how much noise is too much (e.g., noise scale $\sigma\to \infty$), given that legitimate market participants are subject to the same information distortion (Section~\ref{sec:distortion})? 

In what follows, we derive tolerable noise levels that preserve firm profit and consumer welfare with high probability.

We focus on second-order distortions (i.e., profit and welfare fluctuations), not first-order pricing effects. Recall $\Delta$ denote the deviation from the no-noise benchmark (see detailed definitions at Theorem~\ref{thm:pi_dist} and \ref{thm:u_dist}). Formally, given a tolerance $t > 0$, we aim to bound the probability of harmful distortion: $\p(\Delta \leq -t)$:

\begin{lemma}[Profit Distortion Bound] \label{lem:pi_bound}
Fix a distortion tolerance $t_\pi > 0$ and confidence level $1 - \delta \in (0,1)$.
Then under Laplacian noise of scale $\sigma$, the probability that firm $i$'s profit drops below $-t_\pi$ is bounded by
\begin{align}
    \p(\Delta \pi_i \leq -t_\pi) \leq \exp\left(-\frac{t_\pi}{K_\pi \sigma}\right).
\end{align}
To guarantee $\Delta \pi_i \geq -t_\pi$ with probability at least $1 - \delta$, it suffices to impose
\begin{align}
    \sigma \leq \frac{t_\pi}{K_\pi (-\log \delta)},
\end{align}
where $K_\pi := \frac{c^2 (2a_i b_j + c a_j)}{(c^2 - 4b_i b_j)^2}$ is a positive constant. 
\end{lemma}

\begin{lemma}[Welfare Distortion Bound] \label{lem:u_bound}
Fix a welfare distortion threshold $t_U > 0$.
Then the probability that consumer utility drops below $-t_U$ satisfies
\begin{align}
    \p(\Delta U_i \leq -t_U) \lesssim \exp\left(-\frac{t_U}{K_U(c + b_i)\sigma}\right)
\end{align}
To ensure $\Delta U_i \geq -t_U$ with high probability $1 - \delta$, we require
\begin{align}
    \sigma \leq \frac{t_U}{K_U(c + b_i)(-\log \delta)}
\end{align}
where $K_U := \frac{c(2a_i b_j + c a_j)}{(c^2 - 4b_i b_j)^2}.
$ is a positive constant. 
\end{lemma}

Together, these bounds yield the following guarantee:

\begin{theorem}[Upper Bound on Regulatory Noise] \label{thm:noise_harm}
To ensure that distortions in firm profit and consumer utility remain within tolerances $t_\pi, t_U > 0$ with probability at least $1 - \delta$, the noise scale $\sigma$ must satisfy
\begin{align}
    \sigma \leq \min\left\{
    \frac{t_\pi}{K_\pi (-\log \delta)},
    \frac{t_U}{K_U (c + b_i)(-\log \delta)}
\right\}
\end{align}
\end{theorem}

This result complements our earlier findings: while noise slows cartel convergence, its magnitude must remain controlled to avoid harming legitimate sellers and consumers. 

\begin{figure}[H]
    \centering
    \includegraphics[width=0.6\textwidth]{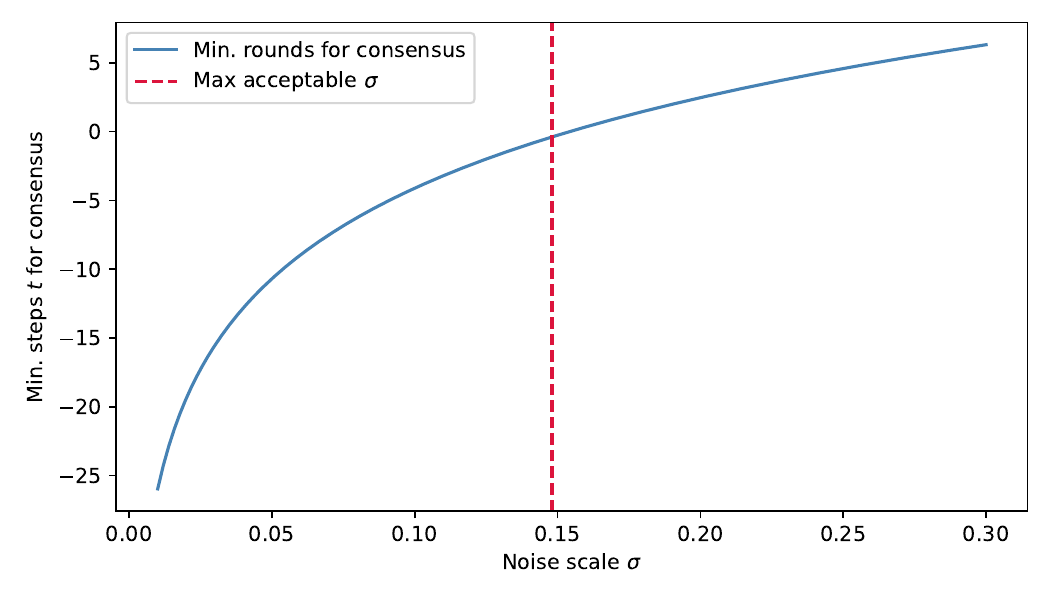}
    \caption{Minimum rounds $t$ required to reach consensus as a function of noise scale $\sigma$. The dashed line marks the maximum acceptable $\sigma$ before distortions to welfare and profit exceed tolerance thresholds with high probability.}
    \label{fig:consensus_delay_bound}
\end{figure}

Figure~\ref{fig:consensus_delay_bound} overlays this constraint as a vertical line over the delay bound (Theorem~\ref{thm:noise_delay}), illustrating how policy makers must balance the trade-off between slowing collusion and avoiding excessive harm to market efficiency.

Theorem~\ref{thm:noise_harm}, thus, defines a viable policy range—strong enough to deter collusion, yet safe enough to preserve market efficiency.

More broadly, our results offer a practical proof of concept: even a heuristic as simple as injecting Laplacian noise into raw market data, thereby limiting information quality, can strike a delicate balance between slowing cartel formation and safeguarding legitimate participants. This highlights how lightweight regulatory levers can achieve strong theoretical guarantees with minimal implementation complexity.

\section{Discussion}

This paper explores whether calibrated information frictions can disrupt algorithmic price coordination. We approached the question in three layers. First, we showed how limited information distorts pricing equilibria, with consequences for profit and consumer welfare. Second, we modeled cartel formation as a belief-averaging process in which forceful agents naturally emerge as leaders. Third, we studied controlled noise as a regulatory lever, showing that it slows cartel convergence while preserving competitive value when calibrated properly.

Our contribution is twofold. Existing work has largely established that algorithms can and do collude, both in simulations and in the field. Our addition is an analytical perspective that views cartel formation through a social learning lens, clarifying not only the mechanics of consensus but also the incentives of leaders and followers and the asymmetric welfare consequences for consumers. On the regulatory side, most prior proposals remain ex post—audits, statistical tests, or forensic detection—whereas our framework highlights an ex ante intervention: shaping the quality of information pricing algorithms receive. This distinction shifts the focus from remedy after collusion to prevention before it takes hold.

The theoretical implications are straightforward. Collusion is not a uniform outcome: it benefits leaders, harms followers, and redistributes consumer welfare asymmetrically. Contrary to common intuition, consumers attached to high-price leaders may even benefit from rivals’ collusion, while followers and their customers lose. Noise provides a simple but sharp policy lever, but its effect depends on calibration: too weak and collusion proceeds, too strong and legitimate firms and consumers are harmed. The feasibility bounds we derive trace out where this balance can be struck.

For managers and policymakers, the message is equally clear. Current law requires explicit proof of communication; in its absence, even suspicious price patterns escape liability. Our results suggest that regulating information inputs offers a lighter-weight, forward-looking safeguard. Implementation is straightforward: rather than redesigning markets or mandating full transparency, regulators can constrain the fidelity of data fed into pricing algorithms. This is particularly timely as AI-driven systems like LLM pricing agents become more prevalent. Future work can test these ideas in richer simulations and practical deployments, but the core message stands: controlled information frictions are a credible, scalable tool for blunting algorithmic collusion.

\bibliographystyle{plainnat}
\begin{spacing}{1.5}
\bibliography{references}    
\end{spacing}

\appendix
\newpage
\begin{center}
    \LARGE{Appendix}
\end{center}

\section{Proofs} \label{app:proofs}
This section contains the proofs for the manuscript. 

\begin{proof}[\bf Proof of Lemma~\ref{lem:opt_p_sym}]
Assuming zero production cost, $k_i=0$, then the objective function becomes:
    \begin{align}
    \pi_i(p_i, p_j) = p_i \cdot D_i(p_i, p_j) = a_i p_i - b_i p_i^2 + c p_i p_j.
    \end{align}
Take the first and second-order derivatives of the objective function: 
\begin{align}
    \text{FOC} &\iff \frac{\partial}{\partial p_i} \pi_i(p_i, p_j) = a_i-2 b_i p_i+c p_j = 0 \\
    \text{SOC Concavity} &\iff \frac{\partial^2}{\partial p_i^2} \pi_i(p_i, p_j) = -2b_i \leq 0
\end{align}
Recall $b_i > 0$, then concavity is trivially satisfied. We may conclude $F_i$'s best response is given by solving first-order condition of $\pi_i$:
    \begin{align}
    p_i^* = \frac{a_i + c p_j}{2 b_i}. \label{eqn:base_opt_res}
    \end{align}
At Nash equilibrium, both firms simultaneously play their best responses, that is
\begin{align}\label{eqn:sym_fixed_sys}
    \begin{cases}
        p_i^* = \frac{a_i + c p_j^*}{2 b_i} \\
        p_j^* = \frac{a_j + c p_i^*}{2 b_j} \\
    \end{cases}
\end{align}
Solving the fixed point system \eqref{eqn:sym_fixed_sys} yields the Nash equilibrium prices for each firm $i,j\in \inset{1,2}$:
    \begin{align}
    p_i^N = \frac{2 a_i b_j + c a_j}{4 b_i b_j - c^2},\quad
    p_j^N = \frac{2 a_j b_i + c a_i}{4 b_i b_j - c^2}, 
    \end{align}

Additionally, notice that the demand function $D_i\inparen{p_i, p_j} \geq 0$ iff $p_i \leq \frac{a_i + cp_j}{b_i}$. Structurally, under perfect information, the best-response mapping automatically rules out negative demand, so all equilibrium outcomes are in the economically relevant region. 
\end{proof}

\begin{proof}[\bf Proof of Lemma~\ref{lem:opt_p_asym}]
    We assume strategic symmetry and rationality: each firm expects its competitor to respond optimally using the best-response rule from the complete information setting (see Equation~\ref{eqn:base_opt_res}), but evaluated under estimated values.
    
    Firm $F_i$ knows $\inset{a_i,b_i,b_j, c}$, but only has an estimate $\hat{a}_j$ of the rival $F_j$'s baseline demand $a_j$.
    
    In particular, $F_i$ can only infer the competitor $F_j$'s best response $p_j^*$ via its estimate $\hat{a}_j$. From $F_i$'s perspective, its own pricing decision $p_i^*$ is known and fixed. $F_i$ assumes $F_j$ uses this $\hat{p}_i = p_i^*$ when computing its own price. This assumption—that each firm’s expectation about the other’s price is ultimately correct—is a useful simplification that allows us to solve for equilibrium.

    Formally, the four equilibrium conditions are:
    \begin{enumerate}[(a)]
        \item $F_i$ maximizes profit against its belief $\hat{p}_j$: $p_i^* = \argmax_{p_i} \; \pi_i(p_i, \hat{p}_j)$, 
        \item $F_j$ maximizes profit against its belief $\hat{p}_i$: $p_j^* = \argmax_{p_j} \; \pi_j(p_j, \hat{p}_i)$, 
        \item $F_i$'s belief is correct in equilibrium: $\hat{p}_j = p_j^*$, and 
        \item $F_j$'s belief is correct in equilibrium: $\hat{p}_i = p_i^*$. 
    \end{enumerate}
    
    From $F_i$'s perspective, its own action $p_i^*$ is fixed. It anticipates $F_j$ will apply the best-response rule from \eqref{eqn:base_opt_res}, but with estimated parameter $\hat{a}_j$:
    \begin{align}
        \hat{p}_j^{*} = \frac{\hat{a}_j + cp_i^*}{2b_j}
    \end{align}
    Using this anticipated response, $F_i$ sets its own price as:
    \begin{align}
        p_i^* = \frac{a_i+c \hat{p}_j^*}{2 b_i} = \frac{a_i+c \cdot \inparen{\frac{\hat{a}_j + cp_i^*}{2b_j}}}{2b_i} \implies p_i^* = \frac{2 a_i b_j+c \hat{a}_j}{4 b_i b_j-c^2}
    \end{align}
    By symmetry, we obtain the Bayesian Nash equilibrium prices:
    \begin{align} 
        p_i^{\BNE} = \frac{2 a_i b_j+c \hat{a}_j}{4 b_i b_j-c^2}, \quad
        p_j^{\BNE} = \frac{2 a_j b_i+c \hat{a}_i}{4 b_i b_j-c^2}. 
    \end{align}
    These prices jointly satisfy all four equilibrium conditions and define the unique Bayesian Nash equilibrium under fixed (but possibly biased or noisy) estimates.
\end{proof}

\begin{proof}[\bf Proof of Theorem~\ref{thm:p_dist}]
    By direct substitution from Lemmas~\ref{lem:opt_p_sym} and \ref{lem:opt_p_asym}, we have
\begin{align}
    \Delta p_i := p_i^{\BNE} - p_i^N = \frac{2 a_i b_j + c \hat{a}_j}{4 b_i b_j - c^2} - \frac{2 a_i b_j + c a_j}{4 b_i b_j - c^2} = \frac{c}{4 b_i b_j - c^2} \cdot (\hat{a}_j - a_j).
\end{align}

Since \( 4 b_i b_j > c^2 \) under the linear demand assumptions, the denominator is strictly positive. Therefore, the sign of \( \Delta p_i \) is the same as the sign of the estimation error \( \hat{a}_j - a_j \). That is,
$ \Delta p_i > 0$ if and only if $\hat{a}_j > a_j$, and \( \Delta p_i < 0 \) if and only if \( \hat{a}_j < a_j \).

The absolute distortion is
\begin{align}
    |\Delta p_i| = \frac{c}{4 b_i b_j - c^2} \cdot |\hat{a}_j - a_j|,
\end{align}
which scales linearly in the size of the estimation error. The positive factor \( \frac{c}{4 b_i b_j - c^2} \) acts as a sensitivity parameter translating belief errors into pricing errors.

\textbf{Comparative statics:} 
Observe that the coefficient \( \kappa(c) := \frac{c}{4 b_i b_j - c^2} \) is increasing in \( c \) on the interval \( (0, \min\{b_i, b_j\}) \). The derivative is
\[
\kappa'(c) = \frac{4 b_i b_j+c^2}{\left(c^2-4 b_i b_j\right){}^2} > 0,
\]
so higher cross-price sensitivity \( c \) amplifies the pricing distortion for any given estimation error.
\end{proof}

\begin{proof}[\bf Proof of Theorem~\ref{thm:pi_dist}]
We compare firm $F_i$'s profit under imperfect versus perfect information. The exact
distortion is
\begin{align}
    \Delta \pi_i:=\pi_i^{\BNE}-\pi_i^{N}
=\pi_i\!\big(p_i^{\BNE},p_j^{\BNE}\big)-\pi_i\!\big(p_i^{N},p_j^{N}\big),
\end{align}
for which a closed form exists but is not transparent.

To obtain an interpretable approximation, apply a first–order Taylor expansion of
$\pi_i(p_i,p_j)=p_i(a_i-b_i p_i + c p_j)$ around the complete-information equilibrium
$(p_i^{N},p_j^{N})$:
\begin{align}
    \Delta\pi_i \approx
\frac{\partial \pi_i}{\partial p_i}\Big|_{(p_i^N,p_j^N)} \Delta p_i
+ \frac{\partial \pi_i}{\partial p_j}\Big|_{(p_i^N,p_j^N)} \Delta p_j.
\end{align}
At the complete-information Nash, the FOC gives
$\tfrac{\partial \pi_i}{\partial p_i}=a_i-2b_i p_i + c p_j = 0$, hence
\begin{align}
    \Delta\pi_i \approx \underbrace{\frac{\partial \pi_i}{\partial p_j}\Big|_{(p_i^N,p_j^N)}}_{=\,c\,p_i^N}\,\Delta p_j
= c\,p_i^N\,\Delta p_j.
\end{align}
From the BNE price formulas,
$\Delta p_j = p_j^{\BNE}-p_j^N = \tfrac{c}{\Delta}\,(\hat a_i - a_i)$ with
$\Delta:=4b_i b_j - c^2>0$.
Therefore
\begin{align}
    \Delta\pi_i \approx \frac{c^2\,p_i^N}{\Delta}\,(\hat a_i - a_i)
= \frac{c^2\,(2 a_i b_j + c a_j)}{(4 b_i b_j - c^2)^2}\,(\hat a_i - a_i).
\end{align}
Since all parameters are positive and $4b_i b_j>c^2$, the coefficient is strictly positive.
Hence $\Delta\pi_i>0$ iff $\hat a_i>a_i$, and $\Delta\pi_i<0$ iff $\hat a_i<a_i$.
Moreover,
\[
\big|\Delta\pi_i\big| \approx \frac{c^2\,(2 a_i b_j + c a_j)}{(4 b_i b_j - c^2)^2}\,\big|\hat a_i - a_i\big|.
\]

\noindent\textbf{Comparative statics:}
Define $f(c):=\frac{c^2\,(2 a_i b_j + c a_j)}{(4 b_i b_j - c^2)^2}$.
A direct derivative yields
\[
f'(c)
= \frac{c\left(4c^2(2a_i b_j + a_j c) + (4 a_i b_j + 3 a_j c)(4 b_i b_j - c^2)\right)}
{(4 b_i b_j - c^2)^3}
> 0,
\]
so larger cross-price sensitivity $c$ amplifies the profit distortion.

This linearization is valid under Interior Feasibility (Definition~\ref{def:interior}) and for moderate belief noise;
if a realization hits a boundary (exit/zero demand), the first-order approximation
should be replaced by the boundary expressions in Appendix~\ref{app:cornercases}.
\end{proof}

\begin{proof}[\bf Proof of Theorem~\ref{thm:u_dist}]
    We begin by expressing consumer welfare distortion as the difference between welfare under imperfect and perfect information:
\begin{align}
    \Delta U_i = U_i^{\BNE} - U_i^{N} = \frac{\left(a_i \left(c^2-2 b_i b_j\right)+c \left(\left(\hat{a}_j-2 a_j\right) b_i-c \hat{a}_i\right)\right){}^2-b_i^2 \left(2 a_i b_j+c a_j\right){}^2}{2 b_i \left(c^2-4 b_i b_j\right){}^2}
\end{align}
As the above expression is not exactly tractable for further analysis, we applied the standard technique of first-order Taylor approximation on $U_i$ around $\inparen{p_i^N, p_j^N}$ as is Theorem~\ref{thm:p_dist}:
\begin{align}
    \Delta U_i &= U_i\inparen{p_i^{\BNE}, p_j^{\BNE}} - U_i \inparen{p_i^N, p_j^N} \\
    &\approx \frac{\partial U_i}{\partial p_i} \inparen{p_i^N, p_j^N} \cdot \Delta p_i + \frac{\partial U_i}{\partial p_j}\inparen{p_i^N, p_j^N} \cdot \Delta p_j \\
    &= \frac{c \left(2 a_i b_j+c a_j\right)}{\left(c^2-4 b_i b_j\right){}^2} \cdot \left( c \inparen{\hat{a}_i-a_i}-b_i\left(\hat{a}_j-a_j\right)\right) \label{eqn:joint_u_dist}
\end{align}
Since all model parameters $a_i, a_j, b_i, b_j, c$ are strictly positive and $4 b_i b_j > c^2$, the coefficient in front of the distortion term is also strictly positive. Therefore, we may conclude the first result that welfare increases as rival estimation error ($\hat{a}_i - a_i$) goes up, or as our estimation error of rival ($\hat{a}_j - a_j$) goes down. 

\noindent \textbf{Comparative statics}: Lastly, due to the positivity of model parameters, we have:
\begin{align}
    \frac{\partial}{\partial c} \inparen{\frac{c \left(2 a_i b_j+c a_j\right)}{\left(c^2-4 b_i b_j\right){}^2}} = \frac{2 \left(a_j \left(4 c b_i b_j+c^3\right)+a_i b_j \left(4 b_i b_j+3 c^2\right)\right)}{\left(4 b_i b_j-c^2\right){}^3} > 0 \\
    \frac{\partial}{\partial c} \inparen{\frac{c \left(2 a_i b_j+c a_j\right)}{\left(c^2-4 b_i b_j\right){}^2} \cdot c }= \frac{c \left(a_j \left(12 c b_i b_j+c^3\right)+4 a_i b_j \left(4 b_i b_j+c^2\right)\right)}{\left(4 b_i b_j-c^2\right){}^3} > 0 
\end{align}
This confirms that as $c$ increases, the distortion effect is magnified—both through the overall scaling factor and specifically through the rival’s estimation error. 
\end{proof}

\begin{proof}[\bf Proof of Theorem~\ref{thm:collusive_incentive}]
    Suppose both firms agree on a common (supracompetitive) price $p^{\text{target}} = \max\inset{p_i, p_j}$ after coordination, where $p_i$ and $p_j$ are firms' initial pricing proposal (see Section~\ref{sec:cartel_consensus} for detailed consensus process) that are interior (Definition~\ref{def:interior}).

    Recall the profit functions (Equation~\ref{eqn:profit}) for firm $F_i, F_j$ are:
    \begin{align}
        \pi_i \inparen{p_i, p_j} &= p_i \cdot D_i \inparen{p_i, p_j} = p_i \left(a_i-b_i p_i+c p_j\right)\\
        \pi_j \inparen{p_i, p_j} &= p_j \cdot D_j \inparen{p_i, p_j} = p_j \left(a_j-b_j p_j+c p_i\right)
    \end{align}

    Then, the profit increase for each firm $F_k$, some $k\in \inset{i,j}$, is:
    \begin{align}
        \Delta \pi_k^{\text{coll}} := \pi_k(p^{\text{target}}, p^{\text{target}}) - \pi_k(p_k, p_{-k})
    \end{align}

    There are two cases to consider:
    \begin{enumerate}[(a)]
        \item Suppose $p^{\text{target}} = p_i$. The profit for each firm becomes:
        \begin{align}
            \pi_i(p^{\text{target}}, p^{\text{target}}) &= \pi_i \inparen{p_i, p_i} = p_i \left(a_i+p_i \left(c-b_i\right)\right)\\
            \pi_j(p^{\text{target}}, p^{\text{target}}) &= \pi_j \inparen{p_i, p_i} = p_i \left(a_j+p_i \left(c-b_j\right)\right)
        \end{align}
        Denote collusive price excess as $\epsilon = p_i - p_j > 0$, then the profit increase is:
        \begin{align}
            \Delta \pi_i^{\text{coll}} &= c p_i \left(p_i-p_j\right) = c \epsilon \inparen{\epsilon + p_j}\\
            \Delta \pi_j^{\text{coll}} &= \left(p_i-p_j\right) \left(a_j+p_i \left(c-b_j\right)-b_j p_j\right) = \epsilon  \left(a_j-b_j \left(2 p_j+\epsilon \right)+c \left(p_j+\epsilon \right)\right)
        \end{align}
        Note that $\Delta \pi_i^{\text{coll}} > 0$ and $\frac{\partial}{\partial \epsilon} \Delta \pi_i^{\text{coll}}= c \left(p_j+2 \epsilon \right) > 0$. That is to say, $F_i$'s strictly benefits from the collusive price, and as the collusive price goes up, it benefits even more.

        While $\Delta \pi_j^{\text{coll}}$ may be positive or negative under Consensus-Interior (Definition~\ref{def:consensus_interior}), 
        if consensus instead drives firm $F_j$ to the boundary, then $\pi_j$ collapses to zero 
        and $\Delta \pi_j^{\text{coll}}<0$ strictly. 
        In either case, $F_i$’s strictly positive profit gain suffices to establish that 
        collusion offers tangible incentive to at least one firm. Hence, there exists 
        nontrivial incentive to coordinate.

        \item Suppose $p^{\text{target}} = p_j$, and denote price excess as $\epsilon = p_j - p_i$. The profit increases are:
        \begin{align}
            \Delta \pi_i^{\text{coll}} &= \left(p_i-p_j\right) \left(-a_i+b_i \left(p_i+p_j\right)-c p_j\right) = \epsilon  \left(a_i-b_i \left(2 p_i+\epsilon \right)+c \left(p_i+\epsilon \right)\right)\\
            \Delta \pi_j^{\text{coll}} &= c p_j \left(p_j-p_i\right) = c \epsilon  \left(p_i+\epsilon \right)
        \end{align}
        This mirrors the earlier case, with $F_j$ now playing the role of the high-price proposer. The analysis proceeds identically and we omit repetition.        
    \end{enumerate}
    In either case, the firm that initially proposed the higher price strictly benefits from convergence to $p^{\text{target}}$. Moreover, the profit gain increases with the price gap $\epsilon$. This establishes that supracompetitive consensus always benefits at least one firm, providing a strict incentive to coordinate.
\end{proof}

\begin{proof}[\bf Proof of Theorem~\ref{thm:collusive_harm}]
    We assume the same setting of pricing as in proof of Theorem~\ref{thm:collusive_incentive}, and denote $\epsilon = \abs{p_i - p_j}$. 

    The welfare increase is defined to be:
    \begin{align}
        \Delta U_k^{\text{coll}} &:= U_k(p^{\text{target}}, p^{\text{target}}) - U_k(p_k, p_{-k}) \\
        &= \frac{1}{2b_k} \cdot \inparen{D_k(p^{\text{target}}, p^{\text{target}})^2 - D_k(p_k, p_{-k})^2} 
    \end{align}
    for the consumers of firm $F_k, k \in \{i, j\}$. Similarly, by symmetry, it suffices to only consider the case that $p^{\text{target}} = p_i$, then
    \begin{align}
        \Delta U_i^{\text{coll}} 
        &= \frac{1}{2b_i} \cdot \inparen{D_i(p_i, p_i)^2 - D_i(p_i, p_j)^2}\\
        &= \frac{1}{2b_i} \cdot \underbrace{\inparen{D_i(p_i, p_i) - D_i(p_i, p_j)}}_{+} \cdot \underbrace{\inparen{D_i(p_i, p_i) + D_i(p_i, p_j)}}_{+} > 0 
    \end{align}
    and 
    \begin{align}
        \Delta U_j^{\text{coll}} 
        &= \frac{1}{2b_j} \cdot \inparen{D_j(p_i, p_i)^2 - D_j(p_i, p_j)^2}\\
        &= \frac{1}{2b_j} \cdot \inparen{D_j(p_i, p_i) - D_j(p_i, p_j)} \cdot \inparen{D_j(p_i, p_i) + D_j(p_i, p_j)} 
    \end{align}
    If consensus-interior (Definition~\ref{def:consensus_interior}) is satisfied, we have
    \begin{align}
        D_j(p_i, p_i) > 0 \implies D_j(p_i, p_i) - D_j(p_i, p_j) < 0 \implies \Delta U_j^{\text{coll}} < 0
    \end{align}
    Otherwise, we have 
    \begin{align}
        D_j(p_i, p_i) = 0 \implies D_j(p_i, p_i) - D_j(p_i, p_j) = - D_j(p_i, p_j) < 0 \implies \Delta U_j^{\text{coll}} < 0
    \end{align}
    Hence, the high-price side's consumers strictly gain $\inparen{\Delta U_i^{\text{coll}} >0}$ while the low-price side’s consumers strictly lose $\inparen{\Delta U_j^{\text{coll}} < 0}$.

\end{proof}

\begin{proof}[\bf Proof of Lemma~\ref{lem:deviation_bound}]
    Notice
    \begin{align}
        x_i(t) - p^{\text{target}} = (1 - \alpha)^t (p_i^{\BNE} - p^{\text{target}})
    \end{align}
    Fixing a tolerance $\epsilon > 0$, we wish bound:
\begin{align}
        &\p\inparen{\abs{x_i(t)-P^{\text{Target}}} \geq \epsilon}\\
        &=\p\inparen{\abs{ p_i^{\BNE} - p^{\text{target}}} \geq \frac{\epsilon}{\inparen{1-\alpha}^t}}\\
        &=\p\inparen{p_j^{\BNE} - p_i^{\BNE} \geq \frac{\epsilon}{\inparen{1-\alpha}^t} \mid p_i^{\BNE} < p_j^{\BNE}}\cdot \p\inparen{p_i^{\BNE} < p_j^{\BNE}}
\end{align}

Assume $A_i \approx A_j$ (i.e., symmetric baseline demand). Notice  $p_j^{\BNE} - p_i^{\BNE} = K(\eta_i - \eta_j)$, then symmetry of zero-mean Laplacian sum implies
\begin{align}
    p_i^{\BNE} < p_j^{\BNE} \iff \eta_i - \eta_j > \frac{A_i - A_j}{K} \approx 0 \Rightarrow \p(p_i^{\BNE} < p_j^{\BNE}) \approx \frac{1}{2} 
\end{align}

Since $\eta_i$'s are sub-exponential with $\norm{\eta_i}_{\psi_1} \sim \sigma$, we have $\norm{\eta_i - \eta_j}_{\psi_1} \leq \norm{\eta_i}_{\psi_1} + \norm{\eta_j}_{\psi_1} \sim \sigma$, where $\psi_1$ denotes the Orlicz sub-exponential norm. A standard sub-exponential tail bound\footnote{See Proposition~2.8.1 and Definition~2.8.4 in \cite{vershynin2025hdp}.} implies:

    \begin{align}
        \p\inparen{\abs{K\inparen{\eta_i - \eta_j}} \geq \frac{\epsilon}{\inparen{1-\alpha}^t} } \lesssim 
        2 \exp\inparen{-\frac{ \epsilon}{(1-\alpha)^t K \sigma}}
    \end{align}
    Chain the results together, we obtain the \emph{deviation} bound:
    \begin{align}
        \p\inparen{\abs{x_i(t) - p^{\text{target}}} \geq \epsilon} = \p\inparen{\abs{ p_i^{\BNE} - p^{\text{target}}} \geq \frac{\epsilon}{\inparen{1-\alpha}^t}} \leq \exp\inparen{-\frac{ \epsilon}{(1-\alpha)^t K \sigma}}
    \end{align}

    In the trivial case where $\sigma = 0$ (i.e., no noise), the bound becomes zero, and consensus is reached deterministically.
\end{proof}

\begin{proof}[\bf Proof of Theorem~\ref{thm:noise_delay}]
    From Lemma~\ref{lem:deviation_bound}, the probability of not achieving $\epsilon$-consensus at round $t$ is bounded above by
\begin{align}
     \p\left(\abs{x_i(t) - P^{\text{target}}} \geq \epsilon \right) \leq \exp\left( - \frac{ \epsilon }{ (1 - \alpha)^t K \sigma } \right)
\end{align}
To ensure this deviation probability is at most $\delta \in \inparen{0,1}$, we require
\begin{align}
    \exp\left( - \frac{ \epsilon }{ (1 - \alpha)^t K \sigma } \right) \leq \delta
\end{align}
Taking logarithms and flip the sign, we get
\begin{align}
    - \frac{ \epsilon }{ (1 - \alpha)^t K \sigma } \leq \log \delta \iff
    \frac{ \epsilon }{ (1 - \alpha)^t K \sigma } \geq -\log\delta
\end{align}
Note $\delta \in (0,1) \implies -\log \delta > 0$, then
\begin{align}
    \inparen{1-\alpha}^t \leq \frac{\epsilon}{K\sigma \cdot \inparen{-\log \delta}}
\end{align}
Recall the fact that $\log$ with base less than 1 is a decreasing function, and notice $1-\alpha \in (0,1)$, then
\begin{align}
    t \geq \log_{1-\alpha}\inparen{\frac{\epsilon}{K\sigma \cdot \inparen{-\log \delta}}}
\end{align}
which proves the result.
\end{proof}

\begin{proof}[\bf Proof of Proposition~\ref{prop:repeat_queries}]
    Each firm observes $n$ i.i.d.\ signals of the form $s_j^{(k)} = a_j + \eta_j^{(k)}$, where $\eta_j \overset{iid}{\sim} \operatorname{Laplace}(\sigma)$.

    Recall the noise has zero mean. It immediately follows that
    \begin{align}
        \E s_j^{(k)} = a_j + \E \eta_j^{(k)} = a_j
    \end{align}
    
    Denote the sample mean estimator as $\bar{s}_j = \frac{1}{n} \sum_{k=1}^n s_j^{(k)}$. By the Law of Large Numbers, $\bar{s}_j$ is \emph{weakly consistent}, i.e., $\bar{s}_j \xrightarrow{p} a_j$. By the Central Limit Theorem, the estimator is asymptotically normal:
\begin{align}
    \sqrt{n}(\bar{s}_j - a_j) \xrightarrow{d} \mathcal{N}(0, 2\sigma^2)
\end{align}
Thus, we know $\bar{s}_i\approx \mathcal{N}\inparen{a_i, \frac{2\sigma^2}{n}}$, which implies repeated queries is still subject to the effect of Laplacian noise, even for large samples. 

We defer rigorous derivation of analogous deviation bounds under Gaussian noise, but note that the qualitative delay effect remains due to persistent variance governed by $\sigma^2/n$.
\end{proof}

\begin{proof}[\bf Proof of Lemma~\ref{lem:pi_bound}]
We bound the probability of profit distortion $\p\inparen{\Delta \pi_i \leq -t_\pi}$. Fix the threshold $t_\pi$, and denote $K_\pi := \frac{c^2 \left(2 a_i b_j+c a_j\right)}{\left(c^2-4 b_i b_j\right){}^2}$. Using exactly the same sub-exponential concentration inequality \citep{vershynin2025hdp} as in the proof of Lemma~\ref{lem:deviation_bound} yields:
\begin{align}
    \p\inparen{\abs{\Delta \pi_i} \geq t_\pi} 
    = \p\inparen{\abs{K_\pi \cdot \eta_i} \geq t_\pi} = \p\inparen{\abs{\eta_i} \geq \frac{t_\pi}{K_\pi}} \lesssim 2 \exp(-\frac{t_\pi}{K_\pi \sigma})
\end{align}

Since we are only interested in controlling the negative effect of the left-tail, we have
\begin{align}
    \p\inparen{\Delta \pi_i \leq -t_\pi} \leq \exp(-\frac{t_\pi}{K_\pi \sigma})
\end{align}
We further bound the expression by $\delta \in (0,1)$ (so that $\log \delta<0$) and solve for $\sigma$:
\begin{align}
    \exp(-\frac{t_\pi}{K_\pi \sigma}) \leq \delta 
    \implies 
    -\frac{t_\pi}{K_\pi \sigma} \leq \log \delta 
    \implies
    \frac{t_\pi}{K_\pi \sigma} \geq -\log \delta 
    \implies
    \sigma \leq \frac{t_\pi}{K_\pi \inparen{-\log \delta}}
\end{align}
The above tells us that to keep profit distortion above $-t_\pi$ with high probability $1-\delta$, the noise scale $\sigma$ must not exceed $\frac{t_\pi}{K_\pi \inparen{-\log \delta}}$.  
\end{proof}

\begin{proof}[\bf Proof of Lemma~\ref{lem:u_bound}]
Similarly, fix the threshold $t_U$, and denote $K_U:= \frac{c \left(2 a_i b_j+c a_j\right)}{\left(c^2-4 b_i b_j\right){}^2}$, then
\begin{align}
    \p\inparen{\abs{\Delta U_i} \geq t_U} = \p\inparen{\abs{K_U \cdot \left( c \cdot \eta_i -b_i \cdot \eta_j\right)} \geq t_U} = \p\inparen{\abs{c \cdot \eta_i -b_i \cdot \eta_j} \geq \frac{t_U}{K_U}}
\end{align}
Apply triangle inequality on Orlicz $\psi_1$-norm yields:
\begin{align}
    \norm{c \cdot \eta_i -b_i \cdot \eta_j}_{\psi_1} \leq c \norm{\eta_i}_{\psi_1} + b_i \norm{\eta_j}_{\psi_1} \sim \inparen{c+b_i} \cdot \sigma 
\end{align}
then sub-exponential inequality tells us: 
\begin{align}
    \p\inparen{\abs{\Delta U_i} \geq t_U} \lesssim 2\exp(-\frac{t_U}{K_U \inparen{c+b_i}\sigma}) \implies \p\inparen{\Delta U_i \leq -t_U} \lesssim \exp(-\frac{t_U}{K_U \inparen{c+b_i}\sigma})
\end{align}
Apply the same high probability bound trick with threshold $\delta$, we obtain 
\begin{align}
    \sigma \leq \frac{t_U}{K_U \inparen{-\log \delta} \inparen{c+b_i}}
\end{align}
To ensure the welfare distortion doesn't drop below $-t_U$ with high probability (at least $1-\delta$), the fluctuation scale $\sigma$ must be small enough.
\end{proof}

\begin{proof}[\bf Proof of Theorem~\ref{thm:noise_harm}]
    The results immediately follows from Lemma~\ref{lem:pi_bound} and \ref{lem:u_bound}. 

    To protect legitimate firm's profit, Lemma~\ref{lem:pi_bound} forces
    \begin{align}
        \sigma \leq \frac{t_\pi}{K_\pi \inparen{-\log \delta}}
    \end{align}

    To protect consumer's welfare, Lemma~\ref{lem:u_bound} forces
    \begin{align}
        \sigma \leq \frac{t_U}{K_U \inparen{-\log \delta} \inparen{c+b_i}}
    \end{align}

    To protect interests of both parties, the inequalities above must both be satisfied, which requires
    \begin{align}
        \sigma \leq \min\left\{
    \frac{t_\pi}{K_\pi (-\log \delta)},
    \frac{t_U}{K_U (c + b_i)(-\log \delta)}
\right\}
    \end{align}
\end{proof}

\section{Derivation of Consumer Welfare} \label{app:consumer_welfare}

Let $p_i^{\max}$ denote the choke price at which demand for firm $F_i$ falls to zero. Solving $D_i = 0$ yields:
\begin{align}
    p_i^{\max} = \frac{a_i + c p_j}{b_i}
\end{align}
The resulting consumer surplus (triangular area) from purchasing firm $F_i$'s product is:
\begin{align}
    U_i(p_i, p_j) &= \frac{1}{2} \cdot D_i(p_i, p_j) \cdot (p_i^{\max} - p_i) \\
    &= \frac{1}{2} \cdot D_i(p_i, p_j) \cdot \frac{D_i(p_i, p_j)}{b_i} \\
    &= \frac{1}{2b_i} \cdot D_i(p_i, p_j)^2 \\
    &= \frac{(a_i - b_i p_i + c p_j)^2}{2b_i} \label{eqn:consumer_welfare}
\end{align}

\section{Boundary Outcomes: Market Exit and Zero Demand}
\label{app:cornercases}

For completeness of Section~\ref{sec:distortion}, we characterize distortions when the baseline pricing equilibrium $p^{\BNE}$ falls outside the interior range $0 < p^{\BNE} < p^{\max}$.

\subsection*{Negative Equilibrium Prices during Market Exit}
\label{app:negative}
When $p^{\BNE} < 0$, truncation $\max\inset{0, p^{\BNE}}$ implies firms post zero prices. We show how the distortion terms collapse in this regime, with profit and consumer welfare both following trivial limits.

Suppose $F_i$ underestimates rival's baseline demand $a_j$ by sufficiently large margins (i.e., $\hat{a}_j \leq -\frac{2 a_i b_j}{c}$), then
\begin{align}
    \frac{2 a_i b_j + c \hat{a}_j }{4 b_i b_j - c^2} < 0 \implies p_i^{\BNE} = 0 
\end{align}

It immediately follows that $\pi_i(p_i^{\BNE}, p_j^{\BNE}) = p_i^{\BNE} \cdot D_i\inparen{p_i^{\BNE}, p_j^{\BNE}} \equiv 0$ for any rival pricing responses $p_j^{\BNE}$.
We may conclude that price and profit distortions are:
\begin{align}
    \Delta p_i =  p_i^{\BNE} - p_i^N &= - p_i^N < 0 \\
    \Delta \pi_i = \pi_i^{\BNE} - \pi_i^N &= - \pi_i^N < 0 
\end{align}
As for consumer welfare, we expand $U_i(p_i,p_j)$ jointly around the Nash benchmark $(p_i^N,p_j^N)$, exactly as in Theorem~\ref{thm:u_dist}. 
Here $\Delta p_i := 0 - p_i^N$ captures the deviation of firm $i$ to zero pricing, and $\Delta p_j := p_j^{\text{BNE}} - p_j^N$ captures the rival’s response. 
The first-order expansion gives
\begin{align}
    \Delta U_i 
    &\approx 
    \underbrace{\frac{\partial U_i}{\partial p_i}\inparen{p_i^N, p_j^N}}_{=-D_i(p_i^N,p_j^N)} \Delta p_i
    + \underbrace{\frac{\partial U_i}{\partial p_j}\inparen{p_i^N, p_j^N}}_{=(c/b_i) D_i(p_i^N,p_j^N)} \Delta p_j \\[6pt]
    &= D_i(p_i^N,p_j^N)\, p_i^N 
    + \frac{c}{b_i}\, D_i(p_i^N,p_j^N)\, \Delta p_j.
\end{align}
Recall demand evaluated at best responses is strictly positive, $D_i(p_i^N,p_j^N) = \frac{b_i \left(2 a_i b_j+c a_j\right)}{4 b_i b_j-c^2} > 0$. Then, we know
\begin{align}
    \Delta U_i \geq 0 \iff p_i^N + \frac{c}{b_i} \cdot \Delta p_j \geq 0 \iff \Delta p_j \geq - \frac{b_i}{c} \cdot p_i^N
\end{align}

\noindent \textbf{Economic Interpretation:}  
The sign of the welfare distortion hinges entirely on how limited information shifts the rival’s pricing. If the rival sets its price above the Nash benchmark, consumers strictly gain. Even if the rival lowers its price, consumer welfare can remain nonnegative provided the cut is modest. 

Strikingly, it is only when the rival becomes too aggressive—reducing its price by more than $-\tfrac{b_i}{c} p_i^N$—that consumers are worse off. Thus, contrary to the conventional view that informational frictions are always harmful to consumers, here excess noise can actually benefit consumers by dampening competition, though extreme rival misperceptions undo this advantage.

\subsection*{Excessive Prices Beyond the Choke Price}
\label{app:excessive}
When $p^{\BNE} > p^{\max}$, demand vanishes. We document the resulting outcomes—zero trade, zero consumer welfare, and profit losses relative to the interior benchmark.

Suppose the posted profile is $\inparen{p_i^{\BNE}, p_j^{\BNE}}$ with $p_i^{\BNE} > p_i^{\max}$. The choke price $p_i^{\max}$ is defined by
\begin{align}
    p_i^{\max} = \frac{a_i + c p_j^{\BNE}}{b_i} \iff D_i\inparen{p_i^{\BNE}, p_j^{\BNE}} = 0
\end{align}
Naturally, without any demand, profit $\pi_i \equiv 0$, and consumer welfare $D_i$ is trivially nil.

Under this regime, the distortions are:
\begin{align}
    \Delta p_i &= p_i^{\BNE} - p_i^N \\
    \Delta \pi &= -\pi_i^N < 0 \\
    \Delta U_i &=  -U_i\inparen{p_i^N, p_j^N} < 0 
\end{align}

\noindent\textbf{Economic Interpretation:}  
To first order, the price distortion $\Delta p_i$ mirrors the structure in Theorem~\ref{thm:p_dist}. At second order, however, both sides of the market are unambiguously worse off: profits collapse to zero, while consumers lose the surplus. This case highlights the perils of excessive information friction—once prices overshoot the choke level, limited information ceases to be benign and instead drives complete market breakdown.

\end{document}